\documentclass[aps,prb,twocolumn,eqsecnum,superscriptaddress,showpacs,10pt]{revtex4-1}
\usepackage{natbib}
\usepackage{mathbbol,amsmath,amsfonts,amssymb,bm,color,dsfont,bbold}
\usepackage{graphicx,psfrag,array,braket,subfig}
\bibpunct{[}{]}{,}{n}{}{}

\def\cT{{\cal T}}

\providecommand{\ignore}[1]{}
\providecommand{\aucmnt}[1]{#1}
\def\be{\begin{equation}}
\def\ee{\end{equation}}
\renewcommand{\aucmnt}[1]{}

\newcommand{\Comment}[1]{}

\newcommand{\Eq}[1]{Eq.~(\ref{#1})}

\begin{document}
\title{Composite fermions in Fock space: Operator algebra, recursion relations, and order parameters}
\author{Li Chen}
\affiliation{National High Magnetic Field Laboratory and Department of Physics, Florida State University, Tallahassee, Florida 32306, USA}
\affiliation{College of Physics and Electronic Science, Hubei Normal University, Huangshi 435002, China}
\author{Sumanta Bandyopadhyay}
\affiliation{Department of Physics, Washington University, St.Louis, Missouri 63130, USA}
\author{Kun Yang}
\affiliation{National High Magnetic Field Laboratory and Department of Physics, Florida State University, Tallahassee, FL 32306, USA}
\author{Alexander Seidel}\affiliation{Department of Physics, Washington University, St.Louis, MO 63130, USA}
\date{\today}
\begin{abstract}
We develop recursion relations, in particle number, for all  (unprojected) Jain  composite fermion (CF) wave functions.
These recursions generalize a similar recursion originally written down by Read for Laughlin states, in mixed first/second-quantized notation.
In contrast, our approach is purely second-quantized, giving rise to an algebraic, ``pure guiding center'' definition of CF states that de-emphasizes first-quantized many-body wave functions. Key to the construction is a second-quantized representation of the flux attachment operator that maps any given fermion state to its CF counterpart.
An algebra of generators of edge excitations is identified. In particular, in those cases where a well-studied parent Hamiltonian exists, its properties can be entirely understood
in the present framework, and the identification of edge state generators can be understood as an instance of ``microscopic bosonization''.
The intimate connection of Read's original recursion with ``non-local order parameters'' generalizes to the present situation, and we are able to give explicit second-quantized
formulas for non-local order parameters associated with CF states.
\end{abstract}
\maketitle
\section{Introduction}

Several years after Laughlin's seminal wave function \cite{Laughlin83} and subsequent hierarchical constructions \cite{haldane_hierarchy, halperin_hierarchy} opened the door for a theoretical understanding of the fractional quantum Hall (FQH) effect, Jain discovered what can be understood as the essential weakly interacting degrees of freedom under a large variety of circumstances: the concept of a ``composite fermion'' \cite{Jain1,Jain3} with no/residual interactions \cite{FQHCF1,FQHCF2,FQHCF3}
offers a compelling way to predict almost all of the  observed plateaus in the lowest Landau level (LL), and has served as the basis for important field theoretical developments \cite{lopez}.
At the same time,
reservations regarding the very definition of a composite fermion have been voiced \cite{dyakonov}.
At a technical level, it is usually defined as a prescription for the construction of variational states via flux attachment, or interchangeably, the attachment of ``zeros'' or ``vortices.'' This somewhat operational definition is of course in good keeping with  tradition in the theory of the FQH effect, which is to give prescriptions for the variational construction of first-quantized many-body wave functions.

In this work, we wish to depart from this tradition. There is an alternative school of thought in the microscopic study of FQH systems that runs counter to the idea of describing states through analytic functions of coordinates.
In a strong magnetic field, activation of degrees of freedom associated with dynamical momenta is energetically costly. These degrees of freedom can therefore be considered either completely or largely frozen out. The physics thus takes place in a reduced Hilbert space that is, to large degree or entirely, stripped of dynamical momenta. This reduced Hilbert space is too coarse to allow for the concept of a position with continuous spectrum. As the orbitals in any given LL may be labeled by an integer, this Hilbert space is properly thought of as associated to a one-dimensional lattice, with the retained LLs (if more than one) interpreted as an internal degree of freedom.
The particle position, once projected onto this Hilbert space, is an operator associated to the classical guiding center coordinate (at least in the limit of only the lowest LL kept). Its components have non-trivial commutation relations. Some approaches to the physics in the FQH regime eschew the use of analytic wave functions in favor of working directly with the algebra of these guiding-center coordinates (e.g., Refs. \onlinecite{Kivelson_PRB_36_1620, Murthy, Haldane2011}).
Here, we wish to give a characterization of the concept of a composite fermion using such an approach.

In the past, we have found fruitful a setting in part motivated by certain varieties of multilayer graphene \cite{McCann06, Barlas12,Jain221}, and in part by an interesting class of parent Hamiltonians \cite{JKT_PRL_64_1297, ReMac, Jainpaper}. Here, the  $n$ lowest Landau levels are  taken to be degenerate, and a local interaction is imposed to stabilize model FQH wave functions within the degenerate LL subspace.
These interacting Hamiltonians
are extraordinary in that there exists a scheme to infer the long distance physics of the state that is both compelling and simple, and leaves very little room for ambiguity.
They unambiguously define a ``zero mode space'' of elementary excitations, and the counting of such zero modes at given angular momentum (relative to the ground state) tends to exactly match the mode counting in a conformal edge theory. Whenever this applies, we will say that the FQH parent Hamiltonian in question satisfies the ``zero mode paradigm.''

There are, of course, many such interesting Hamiltonians known  for the case $n=1$ \cite{RRcount, ardonne,ArdonneJPhysA35:447, SY2}. In this case, some of us have recently shown \cite{ortiz, Chen2014, Mazaheri14}  (for the simplest, Laughlin state parent Hamiltonians) that the rigorous characterization of the zero mode space may be done in a purely  second-quantized framework that does not reference the analytic polynomial wave functions of the traditional approach at all. These developments were very useful for the study of the case $n>1$, which
we argued \cite{Chen17, sumanta18} is absolutely necessary to consider if we wish to establish a zero mode paradigm for a more representative set of FQH states. Indeed, already for the phases described by Jain states, which are elemental to both theory and experiment, we argued that a zero mode paradigm requires $n>1$, with the $n=2$ case extensively discussed in Ref. \onlinecite{Chen17}, where a suitable amalgam of first- and second-quantized methods was adopted. The utility of this mixed first/second-quantized formalism was further demonstrated in Ref. \onlinecite{sumanta18}, where the notion of an ``entangled Pauli principle'' was introduced, and much of the low-energy physics of the non-Abelian, mixed-Landau-level Jain 221 state \cite{Jain2, Jain3, Jain221} was linked to properties of its microscopic two-body parent Hamiltonian.

Here we want to again adopt the ``purist'' point of view of Refs. \onlinecite{ortiz, Chen2014, Mazaheri14}
and describe unprojected, i.e., mixed-Landau-level composite fermion states in a purely algebraic, second-quantized framework, using operators that are all manifestly expressible in terms of electron creation/annihilation operators referring to some preferred LL basis.
In particular, we wish to give such an algebraic definition to the composite fermion and the associated ``vortex-attachment'' concept itself, within the unprojected, mixed-LL setting described above. In doing so, we simultaneously make contact with second-quantized recursion formulas, in particle number $N$, for  Jain-type composite fermion states. This again naturally extends work by some of us on the Laughlin state(s) \cite{ortiz, Chen2014, Mazaheri14}. Moreover, such recursion formulas turn out to be intimately tied to the concept of an order parameter as discussed by
Read \cite{readOP}. This has the added benefit that we are able to give explicit second-quantized formulas for such order parameters.

The remainder of this paper is organized as follows.
In Sec. \ref{n1} we develop the backbone of the formalism for the lowest Landau level only. We develop a second-quantized algebra closely related to symmetric polynomials, following earlier work. We use this to define the composite fermion ``flux-attachment'' operator in second-quantization, \Eq{JLL}. In particular, we develop a recursion relation for this operator that also clarifies its relation to power-sum, or alternatively, elementary symmetric polynomials. From this we rederive a recursion relation \cite{Chen2014} for the Laughlin state. In Sec. \ref{recur}, we generalize all these results to a general number of Landau levels, resulting, in particular, in recursion relations for the second-quantized Jain composite fermion states, Eqs. \eqref{tcpsi1}-\eqref{psi_numberop}. The algebra of symmetric polynomials is replaced with a larger algebra of so-called ``zero-mode generators,'' Eqs. \eqref{Pab}-\eqref{gng}. In Sec. \ref{25rec}, we specialize to two Landau levels, simplifying some of the more general results. In Sect. \ref{PZMP} and \ref{mbos}, we elaborate on implications of our results in the presence of special parent Hamiltonians and in particular for the zero mode structure of such Hamiltonians. For the special case of filling factor $2/5$, some earlier conjectures are proven. The algebra of zero-mode generators, while arising in a microscopic setting,  is shown to encode the effective edge theory.
In Sec. \ref{OPsec}, we use the formalism constructed in preceding sections to give explicit expressions for  $n$-component non-local order parameters for composite fermion states discussed by Read \cite{ReadOP2} in terms of the microscopic electron operators [Eqs. \eqref{OP4} and\eqref{OP5}]. We conclude in Sec. \ref{conclusion}.

\section{Derivation of recursive formula in the lowest Landau level\label{n1}}

The heart of this paper will be a second-quantized formula, recursive in particle number $N$, for the composite fermion vortex attachment operator
\be\label{JN}\begin{split}
\hat J_N:  & \psi(z_1, \bar z_1,\dotsc, z_N, \bar z_N) \\ & \rightarrow\mathfrak{N} \prod\limits_{1\leq i<j\leq N} (z_i-z_j)^M \psi(z_1, \bar z_1,\dotsc, z_N, \bar z_N),
\end{split}
\ee
where $M$ is an even number that we will usually leave implicit,  the $z_i$ are the particle's complex coordinates, and we leave room for a ($N$-dependent) normalization factor $\mathfrak{N}$  that we will not be interested in.
For pedagogical reasons, we will begin our discussion by focusing on the lowest LL ($n=1$) in this section.
A second-quantized recursion relation for the Laughlin state was given earlier in Ref. \onlinecite{Chen2014}. The main difference
between the latter and the developments in this section will be that here we establish the recursion {\em directly} for the
Jastrow vortex-attachment operator $\hat J_N$ itself. This will descend to the earlier recursion for the Laughlin state. However, the extension to the operator $\hat J_N$ will prove essential to the generalization of the recursion formulas to unprojected Jain states (the case $n> 1$).

Considering for now $n=1$, recall that the $N$-particle Laughlin state may be written as
\be
\ket{\psi_N}=\hat J_N\ket{\Omega_N},
\ee
where $\ket{\Omega_N}=c_0^\dag c_1^\dag c_2^\dag\cdots c_{N-1}^\dag \ket{0}$ is an integer quantum Hall state for fermions,
  and the Bose-Einstein condensate $\ket{\Omega_N}=(c_0^\dag)^N \ket{0}$ for bosons,
we will see that a recursion of $\hat J_N$
will descend to a recursion of the Laughlin state. (Here, $\ket{0}$ denotes the vacuum state.) Analogous statements will be true for $n>1$ (Jain states).

The object $\hat J_N$ in \Eq{JN},  can be interchangeably viewed as an operator and as a symmetric polynomial in $N$ variables.
As such, it can be written as $J_N(z_1,\dotsc, z_N)$ or $J_N(p_1,\dotsc, p_N)$ (we will stick to the latter), where the $p_k=\sum_{i=1}^N z_i^k$ are power-sum symmetric polynomials. As a by-product, we will clarify the relation between $J_N$ and such power-sum symmetric polynomials, again via recursion. At operator level, we may then also write
\be \label{JNpoly}
   \hat J_N=J_N(\hat p_1,...,\hat p_N),
\ee
where the $\hat p_k$ are operator representations of the $p_k$ that facilitate the multiplication of first-quantized wave functions with the symmetric polynomial $p_k$. Such representations have been discussed at some length in Refs. \onlinecite{ortiz, Chen2014, Mazaheri14}. They depend slightly on the geometry (and LL basis), where, with the conventions of the ``thick cylinder'', one simply has
\be\label{pk}
 \hat p_k=\sum_m c_{m+k}^\dag c_m \qquad (k\geq 0).
\ee
In this section, we will adopt these thick cylinder conventions for simplicity.
Other geometries differ from the above only by normalization conventions that can be implemented via the replacements
\be\label{geom}
    c_m\rightarrow {\cal N}_m c_m\;,\quad c_m^\dag\rightarrow {\cal N}_m^{-1} c_m^\dag\;,
\ee
which can be facilitated via the similarity transformation $D^{-1} (\;)D$, $D=\exp(\sum_m \ln({\cal N}_m) c^\dag_m c_m)$.
We give the normalization constants ${\cal N}_m$ for various relevant geometries in Table \ref{norm}.
The electron creation/annihilation operators $c_m^\dagger$, $c_m$ refer to lowest LL orbitals with angular momentum $m$ about
the quantization axis. The results in this section will be stated in a manner that is valid for both bosonic as well as fermionic commutation relations, except where explicitly stated otherwise. The sum in \Eq{pk} is generally unrestricted, but we will use the convention $c^\dag_m=c_m=0$ for $m<0$ for the cylinder and disk geometry (thus rendering the cylinder ``half-infinite''), and analogous appropriate restrictions for the sphere.
It should be emphasized that for many of our purposes, the ``first-quantized'' interpretation of the operators
\Eq{pk} as power-sum symmetric polynomials does not matter, but indeed the definition \eqref{pk}
and the resulting algebraic properties are all that we need. For example, it is trivial to verify that the operators
\eqref{pk} all commute ($k\geq0$ !). However, whenever definiteness is required, the term ``symmetric polynomial''
means a polynomial in the complex coordinates $z_i$ only in disk geometry. On the cylinder, it means a polynomial
in the quantities $\xi_i=\exp(\kappa z_i)$, where $\kappa$ is the inverse radius of the cylinder. Analogous statements can be
made for the spherical geometry, which we will not use explicitly in this work, but refer the reader to Ref. \onlinecite{ortiz} for further details in this context. The reader who wishes to focus on the disk should always have the substitutions \eqref{geom} in mind, which do not affect any of the following algebra.

According to a well-known theorem in algebra, any symmetric polynomial ${\cal P}(z_1\dotsc z_N)$ in $N$ variables can be uniquely expressed through a polynomial in $p_1,\dotsc,  p_N$ (${\cal P}=\mathbf { P}(p_1,\dotsc,p_N)$). This includes the $p_k$ for $k>N$. Note, however, that
the operators \Eq{pk} are defined for {\em any } particle number $N$, and the aforementioned polynomial relations between
the $p_{k\in\{1\dotsc N\}}$ and the $p_{k>N}$ carry over to the $\hat p_k$ only  within subspaces of particle number $\leq N$.
Similarly, it is convenient to define $\hat p_0=\sum_m c^\dag_m c_m\equiv \hat N$, which, for fixed particle number $N$,
can be viewed as representing a constant (degree zero) polynomial.

Alternatively, any symmetric polynomials in $N$ variables ${\cal P}(z_1\dotsc z_N)$ can be generated from elementary symmetric polynomials $ e_k=\sum_{1\leq i_1<\dotsc<i_k\leq N}  z_{i_1}\cdot\dotsc\cdot z_{i_k}$, $1\leq k\leq N$, i.e., ${\cal P}= \mathfrak{ P}( e_1,\dotsc, e_N)$, with $\mathfrak{ P}$ a polynomial. Again, we may ask what second-quantized operator facilitates  multiplication with  $ e_k$.
These are \cite{Chen2014, Mazaheri14}
\be\label{ek}\begin{split}
&{\hat  e_k} = \frac{1}{{k!}} \sum_{{l_1,...,l_k} } c_{ {l_1} + 1}^\dag c_{ {l_2} + 1}^\dag \cdots c_{ {l_k} + 1}^\dag c_{ {l_k}} \cdots c_{ {l_2}}c_{ {l_1}},\\& \mbox{with}\quad \hat  e_0:=\mathbb{1}\,,
\end{split}
\ee
given here again for the simple thick cylinder conventions, with disk conventions as detailed in \Eq{geom} and Table \ref{norm}. It is worth noting that unlike the $p_k$, the $ e_k$ vanish automatically for $k>N$. This is respected by the operators $\hat  e_k$, which automatically vanish on any state with particle number $N<k$.
The $\hat  e_k$ and the $\hat p_k$ are related by the Newton-Girard formulas,
\be\label{ngLL} \hat  e_k=\frac{1}{k}\sum\limits_{d=1}^{k}(-1)^{d-1} \hat p_d \hat  e_{k-d}. \ee
These can be directly derived \cite{Mazaheri14} from the operator definitions \eqref{pk} and \eqref{ek}, without any reference to the ``polynomial interpretation'' of these operators. Clearly, \Eq{ngLL} is invariant under the similarity transformation leading to \Eq{geom}, and is thus seen to be geometry independent even if we did not know  about its meaning in terms of polynomials. \Eq{ngLL} may first be used for $k\leq N$ to express all $\hat  e_{k\leq N}$ through $\hat p_{k\leq N}$. Subsequently, letting $\hat  e_{k>N}\equiv 0$, it can be used to explicitly obtain the identities for the $\hat p_{k>N}$
in terms of the $\hat p_{k\leq N}$ mentioned above, valid within the subspace of particle number $\leq N$. Independent of $N$, it is also obvious from these relations that the $\hat  e_k$ commute with one another (as the $\hat p_k$ do), and also commute with all of the $\hat p_k$ (for the same reason).

\begin{table}[t]
\begin{tabular}{ c|c|c|c}
\hline\hline
   & disk  & cylinder & sphere  \\ \hline \\ ${\cal N}_m$&
$\frac{1}{\sqrt{2^m m!}}$  & $\exp(-\frac {1}{2} \kappa^2 m^2)$  & $\frac{1}{(2R)^{m+1}} \sqrt{N_\Phi\choose m}$    \\
\end{tabular}
\caption{Normalization constants ${\cal N}_m$ for various geometries. $ \kappa $ is the inverse radius of the cylinder   $\kappa=1/R_y$. $ R  $ is the radius of the sphere and $N_{\Phi}$ is the number of flux quanta threading the sphere.}
\label{norm}
\end{table}

We will now derive a second-quantized recursive formula for $ \hat J_N $, which turns out to be straightforward to generalize to higher Landau levels.
At the polynomial level, we will also clarify the relation between the Laughlin-Jastrow factor \Eq{JN} and power-sum symmetric polynomials. More precisely, we will give a recursive operator definition
of  $\hat J_N$ {\em both} through electron creation/annihilation operators as well as in terms of polynomial expressions in the $p_k$.

We begin by stating a technical lemma.

{\bf Lemma 0.}
Let $\mathbf{P}(p_0,p_1,\dotsc,p_N)$ be a polynomial in $N+1$ variables. The operator $\mathbf{P}(\hat p_0,\hat p_1,\dotsc,\hat p_N)$
obtained by substituting the operators $\hat p_k$, \Eq{pk}, for $p_k$ satisfies
\be\label{cdagJ}\begin{split} & c_k^\dag \mathbf{P}( \hat p_0, \hat p_1,\dotsc, \hat p_N)\\=&\sum\limits_{l_0,l_1,\dotsc, l_N}\frac{(-1)^{l_0+l_1+\cdots l_N}}{l_0!l_1!\cdots l_N!}\left(\partial^{l_0}_{p_0}\cdots\partial^{l_N}_{p_N}\mathbf{P}\right)( \hat p_0, \hat p_1,\dotsc, \hat p_N)\\& \times c_{k+l_1+2l_2+\cdots+Nl_N}^\dag.\end{split}\ee
Note that we will often be interested only in the action of operators such as $\mathbf P$ within the subspace of fixed particle number $N$. In this context it may not be warranted to have explicit dependence on $\hat p_0$, which is then just a constant, and representing the constant part of $\mathbf{P}$ through $\hat p_0$ may be considered redundant/unnecessary. It is, however, easy to specialize the lemma to the case of no dependence on $\hat p_0$.

{\em Proof of Lemma 0:}
We start by noting
\be \label{cpcomm}
[c_k^\dag,\hat p_r]=-c_{r+k}^\dag,
\ee
trivially obtained from \eqref{pk}, for both fermions and bosons.
 We first prove \Eq{cdagJ} for the case of powers of the form $\mathbf{P}=\hat p_r^d$, by induction in $ d $, then prove the  case of general polynomials by induction in $ N $. For this proof, we will not distinguish between the variables $p_r$ and the operators $\hat p_r$ for notational convenience. Considering now $ \mathbf{P}= p_l^d $, we see that \Eq{cdagJ} is trivially satisfied for $d=0$. Assuming \Eq{cdagJ} is satisfied for $ p_r^{d-1} $, we have
\be \begin{split} & c_k^\dag p_r^{d}\\&=[c_k^\dag,p_r]p_r^{d-1}+p_r(c_k^\dag p_r^{d-1})\\&=-c_{k+r}^\dag p_r^{d-1}+p_r \sum\limits_{l}\frac{(-1)^l}{l!}\left(\partial^{l }_{p_r}p_r^{d-1}\right)c_{k+rl}^\dag\\&= \sum\limits_{l}\frac{(-1)^l}{l!}\left(l\partial^{l-1 }_{p_r}p_r^{d-1}+ p_r\partial^{l }_{p_r}p_r^{d-1}\right)c_{k+rl}^\dag\\ &=\sum\limits_{l}\frac{(-1)^l}{l!}\left(\partial^{l }_{p_r}p_r^{d}\right)c_{k+rl}^\dag,\end{split}\ee
where we used induction in the third and fourth line, and $\partial_x^l x^d = l\partial_x^{l-1}x^{d-1}+x \partial_x^l x^{d-1}$ in the last.
Having proven Eq. \ref{cdagJ} for simple powers of the $p_r$, we now prove it for general polynomials by simple induction in $N$.
By linearity, it is sufficient to consider monomials.
Assume hence that Eq. \ref{cdagJ} is true for $\mathbf{P}=p^{m_{N-1}}_{N-1}\cdots p^{m_{0}}_{0}$. We have
\be\begin{split}& c_k^\dag p^{m_{N}}_{N} p^{m_{N-1}}_{N-1}\cdots p^{m_{0}}_{0} \\
=&
\sum\limits_{l_N}\frac{(-1)^{l_N}}{l_N!}(\partial^{l_N}_{p_N}p_N^{m_N})c_{k+Nl_N}^\dag p^{m_{N-1}}_{N-1}\cdots p^{m_{0}}_{0}\\
=&\sum\limits_{l_N,l_{N-1},\dotsc, l_0 }\frac{(-1)^{l_0+l_1+\cdots l_N}}{l_0!l_1!\cdots l_N!} \\ & \times \left(\partial^{l_0}_{p_0}\cdots\partial^{l_N}_{p_N}p^{m_{N}}_{N} p^{m_{N-1}}_{N-1}\cdots p^{m_{0}}_{0}\right)c_{k+l_1+2l_2+\cdots+Nl_N}^\dag.\end{split}\ee
This concludes our induction proof $\square$.

We now define some useful operators:
\be\label{Sell} \begin{split}
& {\hat  S_\ell } = {( - 1)^\ell }\sum\limits_{{n_1} + {n_2}+\cdots +n_M = \ell } {{\hat  e_{{n_1}}}} {\hat  e_{{n_2}}}\cdots {\hat  e_{{n_M}}}\quad \text{for} \quad \ell\geq0,\\&
\hat  S_\ell=0\quad \text{for} \quad \ell<0.\end{split}\ee
Note that, again, the $\hat  S_\ell$ also depend on $M$, the ``flux attachment'' parameter defined in \Eq{JN}, which we usually leave implicit. With the help of these, we now define the following operator recursion:

\be\label{JLL}\begin{split}
& \hat J_0=\mathbb{1},\\ & \hat J_N=\frac{1}{N}\sum\limits_{r\geq0}\sum\limits_{m\geq0}  c_{ m + r}^\dag \hat  S_{M(N-1)-r}\hat J_{N-1} c_{m }, \end{split}\ee
From this definition, it is not immediately obvious that the operator $\hat J_N$ is of the form
\Eq{JNpoly}, i.e., is a polynomial in the $\hat p_{k\leq N}$.
Our first goal will be to prove precisely that.
This then has two important consequences: 1. Any operator that commutes with all the $\hat p_k$ also commutes with $\hat J_N$ and moreover, 2. the operator $\hat J_N$ acts on $N$-body wave functions via multiplication with a certain symmetric polynomial, since all the $\hat p_k$ have this property.
We will then establish that this polynomial is, up to a normalization, the Laughlin-Jastrow flux-attachment factor, \Eq{JN}.

To see this, we assume $\hat J_{N-1}=J_{N-1}(\hat p_1,\dotsc, \hat p_{N-1})$, $J_{N-1}$ a polynomial. This induction assumption is obviously true for $\hat J_0$. We may then use \Eq{cdagJ} to get the following:

\be\label{JNr} \begin{split}
\hat J_N=&\frac{1}{N}\sum\limits_{r,m}  \sum\limits_{l_1,\dotsc, l_{N-1}}\frac{(-1)^{l_1+\cdots l_{N-1}}}{l_1!\cdots l_{N-1}!}\\& \times\left(\partial^{l_1}_{p_1}\cdots\partial^{l_{N-1}}_{p_{N-1}} S_{M(N-1)-r}J_{N-1}\right)\Bigl|_{p_1\rightarrow\hat p_1,\dotsc} \Bigr. \\&\times c_{m+r+l_1+2l_2+\cdots+(N-1)l_{N-1}}^\dag  c_{m }\\=&\frac{1}{N}\sum\limits_{r}  \sum\limits_{l_1,\dotsc, l_{N-1}}\frac{(-1)^{l_1+\cdots l_{N-1}}}{l_1!\cdots l_{N-1}!}\\& \times\left(\partial^{l_1}_{p_1}\cdots\partial^{l_{N-1}}_{p_{N-1}} S_{M(N-1)-r}J_{N-1}\right)
\Bigl|_{p_1\rightarrow\hat p_1,\dotsc} \Bigr.
\\&\times \hat p_{r+l_1+2l_2+\cdots+(N-1)l_{N-1}}.
\end{split}\ee
In writing the above, $S_\ell$ is a polynomial such that $\hat S_\ell= S_\ell(\hat p_1,\dotsc, \hat p_{N-1})$ {\em when acting on states of $N-1$ particles or less.} We can always achieve this, as explained earlier, by expressing the $\hat  e_{k\leq N-1}$ through the $\hat p_{k\leq N-1}$ in \Eq{Sell}, and letting the $\hat  e_{k\geq N}$ equal to zero.
(Note that if $\hat J_N$ acts on $N$-particle states, then $\hat J_{N-1}$ in \Eq{JLL} acts on $N-1$ particle states.)
We may similarly express all the terminal $\hat p$-operators in the last line of \Eq{JNr} through the $\hat p_{k\leq N}$. With these replacements, the difference between \Eq{JLL} and \Eq{JNr} strictly speaking vanishes only on states with particle number $\leq N$. However, since we will exclusively be interested in the action of $\hat J_N$ on states with $N$ particles, this difference can be ignored in the following. Anticipating that the last two equations really define the composite fermion operator \eqref{JN}, we see that \Eq{JNr}, viewed as an equation for symmetric polynomials (i.e., omitting hats) gives a recursive definition of the (even $M$) Laughlin-Jastrow factor in terms of power-sum symmetric polynomials. In this polynomial sense, \Eq{JNr} must of course be correct independent of the number of LLs kept, {\em unlike} the operator definitions given in this section, which so far stand only for the lowest LL. Working backwards from \Eq{JNr}, we will be able to generalize the operator recursion \eqref{JLL} to higher Landau levels.

Before we do this, we give applications of \Eq{JLL} within the lowest LL, and in doing so, establish correspondence with \Eq{JN}. Consider now fermions and the
$N$-particle  state
\be\label{LS2ndq} \ket{\psi_N}=\hat J_N c_0^\dag c_1^\dag \cdots c_{N-1}^\dag\ket{0}.\ee
We will use \Eq{JN} to re-establish a recursive relation for this state, from which, via Ref. \onlinecite{Chen2014} it is then known that \Eq{LS2ndq} defines the densest zero mode
of a pseudo-potential Hamiltonian (for $M=2$, the $V_1$ Haldane pseudo-potential), thus identifying it uniquely as the $1/(M+1)$ Laughlin state.

 From the definition of $\hat J_N$ in Eq. \ref{JLL}, we can prove the following identity
\be\label{cJ}c_r \hat J_N=\sum\limits_{m}\hat  S_{M(N-1)-r+m}\hat J_{N-1} c_{m }. \ee
 The proof of Eq. \ref{cJ} is given in Appendix \ref{proof}.
Using Eq. \ref{cJ}, we obtain \be\begin{split}
c_r\ket{\psi_N}=&\sum\limits_{m}\hat  S_{M(N-1)-r+m}\hat J_{N-1}(-1)^m \\& \times c_0^\dag \cdots c_{m-1}^\dag c_{m+1}^\dag \cdots c_{N-1}^\dag \ket{0}.\end{split}\ee We observe that $ c_0^\dag \cdots c_{m-1}^\dag c_{m+1}^\dag \cdots c_{N-1}^\dag \ket{0} $ is just \be \hat  e_{N-1-m}c_0^\dag c_1^\dag \cdots c_{N-2}^\dag\ket{0} \ee using the definition of $ \hat  e_k $
in Eq. \ref{ek}. Thus we have
\be\label{hole0} c_r \ket{\psi_N}=\sum\limits_{m}\hat  S_{M(N-1)-r+m}(-1)^m \hat  e_{N-1-m}\ket{\psi_{N-1}}
\ee
in which we have used that $\hat  J_{N-1} $, being a polynomial in the $\hat p_k$, commutes with  $ \hat  e_{N-1-m} $. The latter can be written more suggestively
after defining \begin{equation}\begin{split} &{\hat  S_\ell^\sharp } = {( - 1)^\ell }\sum\limits_{{n_1} + {n_2}+\cdots +n_{M+1} = \ell } {{\hat  e_{{n_1}}}} {\hat  e_{{n_2}}}\cdots {\hat  e_{{n_{M+1}}}} \quad\text{for} \quad\ell\geq0,\\& \hat  S^\sharp_\ell=0 \quad\text{for}\quad\ell<0,\end{split}\end{equation}
i.e., $\hat S^\sharp_\ell$ is defined just as $\hat S_\ell$ but with the odd number $M+1$ replacing the even number $M$. With this we can rewrite \Eq{hole0} as
\be\label{hole} c_r\ket{\psi_N} =(-1)^{N-1}\,\hat  S^\sharp_{(M+1)(N-1)-r}\ket{\psi_{N-1}},
\ee
which, up to a constant $ (-1)^{N-1} $ amounting to a phase convention, is the same   as that obtained in Ref. \onlinecite{Chen2014} for the Laughlin state with filling fraction $ 1/(M+1)$.  This formula and its generalizations will be crucial in much of the following. It should be read as follows: The operator $c_r$ creates a (charge 1) hole of well-defined angular momentum. Due to bulk-edge correspondence, such a hole can always be interpreted as an edge excitation of the $N-1$ particle incompressible state, though possibly one of high energy,  living deeply in the bulk of the system. As we have explained elsewhere \cite{Chen2014,Mazaheri14},  the operator $\hat  S^\sharp_\ell$ and the $\hat  e_k$ which it is composed of should be thought of as generators of such edge excitations when acting on the incompressible state. To make these notions more precise, one may consider a pseudo-potential Hamiltonian of the form \cite{haldane_hierarchy}
\be\label{H}
H=V_1+V_3+\dotsc+V_{M-1},
\ee
where the positive operator $V_k$ is (proportional to) the $k$th Haldane pseudo-potential.
It is well-known that the $1/(M+1)$ Laughlin state is the densest zero energy mode (zero mode) of this Hamiltonian, and one may {\em define} quasi-hole/edge excitations as the set of all other zero modes of the same Hamiltonian.
It is easy to see \cite{Chen2014} that the left-hand side of \Eq{hole} is a zero mode if $\ket{\psi_N}$ is, and the $\hat  e_k$ can be shown \cite{Mazaheri14} to generate a complete set of zero modes of the same particle number when acting on the incompressible $1/(M+1)$ Laughlin state. Equation (\ref{hole}) is the precise way to express the charge-1 quasi-hole $c_r |\psi_N\rangle$ in this manner, i.e., as a superposition of edge excitations created in the state $|\psi_{N-1}\rangle$.

At this point, a recursion for the Laughlin state can be obtained following the logic of Ref. \onlinecite{Chen2014}. Applying the operator $c_r^\dagger$ to \Eq{hole} and summing over $r$ produces a factor of the particle number $N$ on the left-hand side. Dividing by this factor gives
\be \label{Lrec}
\ket{\psi_N} = \frac{1}{N}\sum_r
(-1)^{N-1}\,c^\dagger_r\hat  S^\sharp_{(M+1)(N-1)-r}\ket{\psi_{N-1}}\,.
\ee
This recursion, with $\ket{\psi_1} = c^\dagger_0\ket{0}$, has been shown in Ref. \onlinecite{Chen2014} to give the densest (lowest angular momentum) zero mode of the Hamiltonian \eqref{H}, thus uniquely identifying the $\ket{\psi_N}$, \Eq{LS2ndq}, as the $1/(M+1)$ Laughlin state (defined up to an overall constant). As we have shown above, the effect of the operator $\hat J_N$ on {\em any} $N$-particle state is the multiplication of the state's
wave function with a fixed symmetric polynomial
$J_N(p_1,\dotsc, p_N)$.
We may find this polynomial by looking at
\Eq{LS2ndq}, which we now know to be the Laughlin state.
From this equation,  we thus have
\be
\mathfrak{N} \prod_{i<j}(z_i-z_j)^{M+1} = J_N(p_1,\dotsc, p_N)\,\prod_{i<j} (z_i-z_j)\,,
\ee
where the left-hand side is the $1/(M+1)$ Laughlin state, on the right-hand side we used that
$ c_0^\dag c_1^\dag \cdots c_{N-1}^\dag\ket{0}$ in \Eq{LS2ndq} is just a Vandermonde determinant, and we dropped Gaussian factors on both sides. This determines the polynomial $J_N(p_1,\dotsc, p_N)$ to be the Laughlin-Jastrow factor in \Eq{JN}. The same derivation is possible for bosons with very few changes.

We remark that a variant of the recursion \eqref{Lrec} that uses mixed first/second-quantized notation was first given by Read \cite{readOP} (see also Sec. \ref{OPsec} below).
The operator-level recursion \eqref{JLL} for the composite fermion flux attachment is more general, however, as it implies the Laughlin state recursion, but cannot be derived from the latter.
Moreover, it has more general uses which we will turn to in the following. For one, it immediately gives rise to similar recursions for (unprojected) Jain-type composite fermion states. Moreover, there is a general connection between the recursion for the Laughlin states, and an ``order-parameter'' construction for these states, as discussed by Read \cite{readOP}.
Via \Eq{JLL}, systematic generalization of this connection to Jain states will  be possible.

\section{Derivation of recursive formulas for multiple Landau level composite fermion states}\label{recur}

\subsection{Operator recursion\label{OR}}

In the preceding section  we have constructed a recursion relation for the lowest Landau level composite fermion (Laughlin) state $\ket{\psi_N}$.
The central ingredient was the recursion for the Jastrow (CF flux attachment) operator $\hat J_N$, \Eq{JLL}. The key to the generalization of this recursion to higher-LL CF states is the fact that this recursion is the operator manifestation of a polynomial recursion, which we have formally expressed as \eqref{JNr}. This last equation must remain valid, since in any number of LLs the (M-dependent) Jastrow factor is always represented by the same symmetric polynomial in the holomorphic coordinates. As we emphasized earlier, the second-quantized operators associated to the multiplication with such polynomials somewhat depend on the geometry in question, at least when the standard orbital basis for that geometry is used. At the same time, they depend on the number of Landau levels kept. The goal is now to work out the second-quantized operator equations of the last section for the case of multiple LLs, especially the recursion \Eq{JLL}.
Our strategy will be to work backwards from \Eq{JNr}, which is essentially a statement about polynomials and which therefore holds independent of the number of LLs.
The glue between these two equations was the general \Eq{cdagJ}, which flows from the elementary \Eq{cpcomm}.
We thus begin by re-establishing relations concerning the operators associated with power-sum and elementary symmetric polynomials. We will consider $n=2$ first, from which the general structure will become obvious. In Sec. \ref{n1} we used thick cylinder conventions for pedagogical reasons.
In the presence of multiple Landau levels,
the advantage of this geometry is less immediate, and hence we will start by
working in disk geometry. The following treatment will specialize to a rederivation of most of the results of Sec. \ref{n1} for disk geometry when all the higher LL creation/annihilation operators are set equal to zero.

We start by giving the equation
for the operator $\hat p_k$, which again describes the multiplication with the polynomial $\sum_{i=1}^N z_i^k$. As before, these are single-particle operators, and can be straightforwardly worked out in second-quantization from their first-quantized definition. We quote them from
 Ref. \onlinecite{Chen17}:
\be\label{pdold}\begin{split} \hat p_k  &= \sum_{r=0}^{+\infty} \sqrt{\frac{(r+k)!}{r!}}c^\dagger_{0,r+k}c_{0,r}  +\sum_{r=-1}^{+\infty}k \sqrt{\frac{(r+k)!}{(r+1)!}}c^\dagger_{0,r+k}c_{1,r}\\& + \sum_{r=-1}^{+\infty}\sqrt{\frac{(r+k+1)!}{(r+1)!}}c^\dagger_{1,r+k}c_{1,r}.\end{split}\ee
Here, the operator $c_{m,r}$ now refers to the orbital with angular momentum $r$ in the $m$th LL, with $r\geq -m$. An inconvenience is the fact that the commutator $[c_{m,r}^\dag,\hat p_k]$ is not diagonal in $m$, i.e., in general produces terms referring to Landau levels other than $m$. This precludes straightforward generalization of \Eq{cdagJ} (Lemma 0), which rests on the simple form of \Eq{cpcomm}.
 However, one can rewrite the Eq. \ref{pdold} as

\be\label{pd}\begin{split} \hat p_k&= \sum_{r=0}^{+\infty} \sqrt{\frac{(r+k)!}{r!}}c^\dagger_{0,r+k}(c_{0,r}-\sqrt{r+1}c_{1,r}) \\&+\sum_{r=-1}^{+\infty}\sqrt{\frac{(r+k+1)!}{(r+1)!}}(c^\dagger_{1,r+k}+\sqrt{r+k+1}c^\dagger_{0,r+k})c_{1,r}.\end{split}\ee
It turns out that the operators made explicit in this factorization have favorable commutation relations.
We introduce ``pseudo-fermions''
\be\label{tsum} \tilde{c}^*_{a,r}=\sum_b A(r)_{ab}c^\dagger_{b,r};~~~~\tilde{c}_{a,r}=\sum_b {A(r)}^{-1}_{ba}c_{b,r},\ee where \be\label{A2} A(r)=\begin{pmatrix}
\sqrt{r!}&0\\
\sqrt{(1+r)!(1+r)}&\sqrt{(1+r)!}
\end{pmatrix},\ee
and note that $ \tilde c_{i,r}^*  \ne\tilde c_{i,r}^\dag   $, but we still have anti-commutation relations\be \begin{split}
&\{\tilde c_{i,r},\tilde c_{j,r'}^* \}=\delta_{i,j}\delta_{r,r'}\\&\{\tilde c_{i,r},\tilde c_{j,r'} \}=\{\tilde c_{i,r}^*,\tilde c_{j,r'}^* \}=0.
\end{split}\ee
The restriction $r\geq -i$ of the $c_{i,r}$ and $c_{i,r}^\dagger$-operators carries over to the
$\tilde c_{i,r}$ and $\tilde c_{i,r}^\ast$-operators,
As usual, we will use the convention $\tilde c_{i,r}=\tilde c_{i,r}^\ast=0$ whenever $r$ lies outside this range. The significance of the operators $\tilde c_{i,r}^\ast$ is that they create the non-orthogonal, non-normalized single-particle states $z^{i+r}\bar z^i$ (Gaussians are omitted).
This gives
\be\label{pt} \hat p_k= \sum_{a=0,1}\sum_{r=-a}^{+\infty}\tilde c^*_{a,r+k}\tilde c_{a,r}\ee such that
\be
\label{cspcomm}
[\tilde c_{a,r}^*,\hat p_k]=-\tilde c_{a,r+k}^*,  \ee
which is analogous to \Eq{cpcomm}, with the ``LL level like'' basis label $a$ a pure spectator.
We still have $\hat p_0=\hat N$. Observe that if we specialize to a single LL, the transformation \eqref{tsum} facilitates just the similarity transformation discussed in the preceding section. The only difference is that here we do not view this as an ``active'' transformation between different geometries, but rather as a ``passive'' change of basis, involving a non-orthonormal basis (though still orthogonal for $n=1$).

With this new expression for the $\hat p_k$, it is straightforward to adapt the operators for the elementary symmetric polynomials:

\be\label{e}\begin{split}& {\hat e_k} = \frac{1}{{k!}}\sum_{{a_1,...,a_k} = 0,1} \sum_{l_1,...,l_k} \tilde c_{a_1, {l_1} + 1}^{*} \tilde c_{a_2, {l_2} + 1}^{*} \cdots \tilde  c_{a_k, {l_k} + 1}^{*}\\&\times\tilde c_{{a_k}, {l_k}} \cdots \tilde c_{{a_2}, {l_2}}\tilde c_{{a_1},{l_1}}\\&\text{for}\quad k>0,   \\& \hat e_0=\mathbb{1},\qquad \hat e_k=0 \quad \text{for}\quad k<0.\end{split}\ee
Indeed,
the $\hat e_k$ and $\hat p_k$  still satisfy the Newton-Girard formula Eq. \ref{ngLL}. Given that the $\hat p_k$ represent power-sum symmetric polynomials, this again uniquely identifies the $\hat e_k$ in the above equation as representing elementary symmetric polynomials. Owing to Eqs. \eqref{pt} and \eqref{cspcomm}, the proof that Newton-Girard equations are satisfied is a straightforward generalization of that given in Ref. \onlinecite{Mazaheri14} for the LLL. Details are given in Appendix \ref{z}.

In a similar vein, one then easily generalizes \Eq{cdagJ} to the present situation, using the same procedure as in Sec. \ref{n1}:
\be\label{cdagJn2}\begin{split} & \tilde c_{a,k}^* \mathbf{P}( \hat p_0, \hat p_1,\dotsc, \hat p_N)\\=&\sum\limits_{l_0,l_1,\dotsc, l_N}\frac{(-1)^{l_0+l_1+\cdots l_N}}{l_0!l_1!\cdots l_N!}\left(\partial^{l_0}_{p_0}\cdots\partial^{l_N}_{p_N}\mathbf{P}\right)( \hat p_0, \hat p_1,\dotsc, \hat p_N)\\& \times\tilde c_{a,k+l_1+2l_2+\cdots+Nl_N}^*.\end{split}\ee
 With this it is a simple task to carry out the program described at the beginning of this section:
We take the last line of \Eq{JNr} as the recursive definition of the $\hat J_N$ operator, with $\hat J_0=\mathbb{1}$. From this we easily obtain, using the generalized \Eq{cdagJ}, a generalized version of the operator recursion \eqref{JLL}:

\be\label{Jgen}\begin{split}
 \hat J_0=& \mathbb{1},\\  \hat J_N=&\frac{1}{N}\sum_a\sum\limits_{r\geq0}\sum\limits_{m\geq -a}  \tilde c_{a, m + r}^\ast \hat  S_{M(N-1)-r} \hat J_{N-1} \tilde c_{a, m }, \end{split}\ee

Lastly, just as in the preceding section, and as explained in Appendix \ref{proof}, we obtain from this the generalization of \Eq{cJ}:

\be\label{tcJany}\tilde c_{a,r} \hat J_N=\sum\limits_{m\geq-a}\hat  S_{M(N-1)-r+m}\hat J_{N-1} \tilde c_{a,m}.\ee
With all the key ingredients in hand, let us now construct the densest  composite fermion states occupying two Landau levels, also known as $\Lambda$-levels ($\Lambda$Ls) in this context \cite{jainlambda}.
These are just the Jain states at filling factor $2/(2M+1)$. We define

\be\label{zmpsi}\begin{split}\ket{\psi_{2N}}\sim &\hat J_{2N}c^\dagger_{1,-1}c^\dagger_{1,0}...c^\dagger_{1,N-2}c^\dagger_{0,0}c^\dagger_{0,1}...c^\dagger_{0,N-1}\ket{0},\\\ket{\psi_{2N+1}}\sim &\hat J_{2N+1}c^\dagger_{1,-1}c^\dagger_{1,0}...c^\dagger_{1,N-2}c^\dagger_{0,0}c^\dagger_{0,1}...c^\dagger_{0,N-1}c^\dagger_{0,N}\ket{0}\end{split}\ee
for particle number $ 2N $ and $ 2N+1 $, respectively. It is easy to see that, up to normalization factors, these are exactly equal to \be\label{Jain2LL}\begin{split}\ket{\psi_{2N}}=&\hat J_{2N}\tilde c^{*}_{1,-1}\tilde c^{*}_{1,0}...\tilde c^{*}_{1,N-2}\tilde c^{*}_{0,0}\tilde c^{*}_{0,1}...\tilde c^{*}_{0,N-1}\ket{0},\\\ket{\psi_{2N+1}}=&\hat J_{2N+1}\tilde c^{*}_{1,-1}\tilde c^{*}_{1,0}...\tilde c^{*}_{1,N-2}\tilde c^{*}_{0,0}\tilde c^{*}_{0,1}...\tilde c^{*}_{0,N-1}\tilde c^{*}_{0,N}\ket{0},\end{split}\ee
which we use to fix the normalization.
We note that for $M=2$ this defines precisely the Jain 2/5 state, for which again a local pseudo-potential Hamiltonian can be given, such that the states \eqref{Jain2LL} are densest zero modes \cite{JKT_PRL_64_1297,ReMac, Chen17}.  It can be shown that the set of all ($N$-particle) zero modes of this Hamiltonian is precisely the {\em range} of the operator $\hat J_N$, that is, the set generated from states obtained when $\hat J_N$ acts on general $N$-particle Slater determinants \cite{Chen17}, as opposed to  only the {\em densest} Slater determinants used in the definitions \eqref{Jain2LL}.
For the cases $M>2$ and/or $n>2$, there exist, to our knowledge, no local parent Hamiltonians with similar properties in the literature, and we leave their discussion as an interesting problem for the future. For these cases, we will simply {\em define } the $N$-particle zero mode
space as the range of the operator $\hat J_N$.

\subsection{Zero Mode Generators\label{ZMGs}}
Before we further apply the results of this section,
we need to introduce a larger set of operators that we will think of as ``zero mode generators''. Also, we use this opportunity to generalize the setting of the preceding subsection from $2$ to a general number of $n$ LLs. This is straightforward in principle. Essentially,
all it takes is to generalize
 Eq. \ref{tsum} by means of an appropriate $n\times n$ matrix $A(r)$. The explicit form of $A(r)$ is given in Appendix \ref{AppA}.

 In the following, we will be interested in the generalization of the recursive formulas for the ($n=1$) Laughlin state to the $n$-$\Lambda$L
 composite fermion states, in particular, \eqref{Jain2LL} for $n=2$.
 In addition to the operator recursion \eqref{Jgen}, this requires an understanding of zero mode generators, i.e., operators like the $\hat e_k$ and $\hat p_k$ that generate more (possibly, all) zero modes when acting on the ``incompressible'' (densest) zero mode.  To this end,
 in the $n$ LL system,  one can construct additional operators which will satisfy a modified Newton-Girard formula, namely,
\be\label{Pab}
\hat p^{a,b}_k=\sum_{r=-b}^{+\infty}\tilde{c}^*_{a,r+k}\tilde{c}_{b,r}, \qquad (k\geq b-a)\ee
and, for $a\geq b-1$, $k\geq 0$, \be\label{NG2LL} \hat e^{a,b}_k=\frac{1}{k}\hat p^{a,b}_1\hat e^{a,b}_{k-1}+\frac{\delta_{a,b}}{k}\sum\limits_{d=2}^{k}(-1)^{d-1}\hat p^{a,b}_d \hat e^{a,b}_{k-d},\ee where  $\hat e^{a,b}_k$ can be written explicitly,  \begin{equation}\label{gng}\begin{split}
& \hat e^{a,b}_k =\frac{1}{k!}\sum_{l_1,...,l_k=-b}^{+\infty} \tilde c_{a, {l_1} + 1}^{*} \tilde c_{a, {l_2} + 1}^{*} \cdots\tilde c_{a, {l_k} + 1}^{*}\\&\times\tilde c_{{b}, {l_k}} \cdots \tilde c_{{b}, {l_2}}\tilde c_{{b}, {l_1}}, \qquad (k>0), \\&\hat e^{a,b}_0=\mathbb{1}.
\end{split} \end{equation}

The proof of these (modified) Newton-Girard formulas is given in Appendix \ref{z}.
It is through the introduction of these new operators that our formalism offers true advantage over a first-quantized language of polynomials. Unlike the $\hat p_k$, $\hat e_k$, Eqs. \eqref{Pab}, \eqref{gng} have no particularly natural presentation in polynomial language (see below), but still have the favorable algebraic properties discussed here.


The significance of these operators is the following.
We identify the operators $\hat p^{a,a}_d$
as the operators that send
first-quantized expressions of the form $\bar z ^a z^{a+\ell}$ to $\bar z ^a z^{a+\ell+d}$. It is then clear that the operator
\be \hat p_d=\sum_{a,b} \delta_{a,b} \hat p^{a,b}_d.\ee
multiplies any single-particle wave function by $z^d$, and, in the general many-particle context, can be identified as the operator associated with the power-sum polynomial $p_d$ as before. 
In order to further motivate the physical meaning of $\hat p^{a,b}_d$, let us look into their commutation relations,\be\label{Lie}[\hat p^{a,b}_k,\hat p^{b',a'}_{k'}]=\delta_{b,b'}\hat p^{a,a'}_{k+k'}-\delta_{a,a'}\hat p^{b',b}_{k+k'}. \ee
This immediately implies \be\label{Pp}[\hat p^{a,b}_k, \hat p_{k'}] =0. \ee
For the composite fermion operator $\hat J_N$, on the other hand, we will always use the recursion \Eq{JNr} as the defining property. Therefore, as before, the $\hat J_N$ are always expressible through the $\hat p_k$. The last equation then gives
\be\label{Je}[\hat J_N,\hat p_k^{a,b}]= [\hat J_N,\hat e_k^{a,b}]=0,\ee
where, for the $\hat e^{a,b}_k$, we have used the fact that by the relations \eqref{NG2LL}, we can express all of the latter through the $\hat p^{a,b}_k$.
As explained/defined above, the space of all zero modes is precisely the range of the operator $\hat J_N$. Equations (\ref{Je}) then say that the zero mode space is {\em invariant} under the action of the $\hat p_k^{a,b}$ or $\hat e_k^{a,b}$. That is, when any of these operators acts on a zero mode, a new zero mode results. It is for this reason that we think of these operators as zero mode generators.
It is further true that we can generate any $N$-particle zero modes by repeatedly acting with these generators on certain incompressible zero modes $\psi_N$, such as the Laughlin state or a Jain state. In this sense, it turns out that we can in particular think of the $\hat p_k^{a,b}$  as a complete set of zero mode generators.

 We close this section by remarking that with the generalized
 $A(r)$ matrix of  Appendix \ref{AppA},
 Eqs. \eqref{Jgen} and \eqref{tcJany} generalize without change to $n>2$ LLs.

\subsection{Recursion formulas for general composite fermion states\label{state_recur}}
Let us consider the second-quantized composite fermion wave function at the filling fraction $\nu=\frac{n}{Mn+1}$ for $N=n L_{\text{max}}  +q$ particles with $1\leq q \leq n$,
\be\label{genCFstate}
\ket{\psi_N}=\hat J_N\ket{\Psi_N}.
\ee
Explicitly,
\begin{widetext}
\be\begin{split}\ket{\Psi_{N}}=& \tilde c^{*}_{n-1,-(n-1)}\tilde c^{*}_{n-1,-(n-2)} \cdots\tilde c^{*}_{n-1,L_{\text{max}}-n}\tilde c^{*}_{n-2,-(n-2)}\tilde c^{*}_{n-2,-(n-3)}\cdots\tilde c^{*}_{n-2,L_{\text{max}}-n+1}\cdots \tilde c^{*}_{q,-q}\tilde c^{*}_{q,-q+1}\cdots\tilde c^{*}_{q,L_{\text{max}}-q-1}\\&\times \tilde c^{*}_{q-1,-(q-1)}\cdots\tilde c^{*}_{q-1,L_{\text{max}}-q+1} \cdots\tilde c^{*}_{0,0} \cdots\tilde c^{*}_{0,L_{\text{max}}}  \ket{0}.\end{split}\ee
\end{widetext} By abuse of terminology, we will now refer to the index $r$ in $\tilde c_{r,j}^\ast$ as a $\Lambda$-level  index, and to the orbitals created by $\tilde c_{r,j}^\ast$ with fixed $r$ as a $\Lambda$-level.
The wave function $\ket{\Psi_N}$ corresponds to the state in which   $0,1...q-1$th $\Lambda$-levels  each have $L_{\text{max}}+1$ particles, and   $q,q+1,...,n-1$th $\Lambda$-levels  each have   $L_{\text{max}}$ particles.

\begin{figure*}
\subfloat[]
{\includegraphics[width=0.4\textwidth]{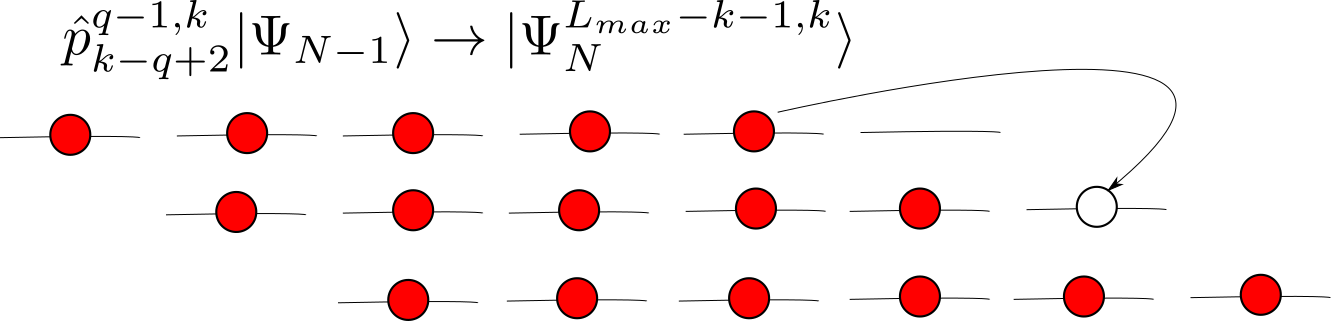}~~~~~~~~~~~}
\subfloat[]{~~~~~~~~~~~\includegraphics[width=0.4\textwidth]{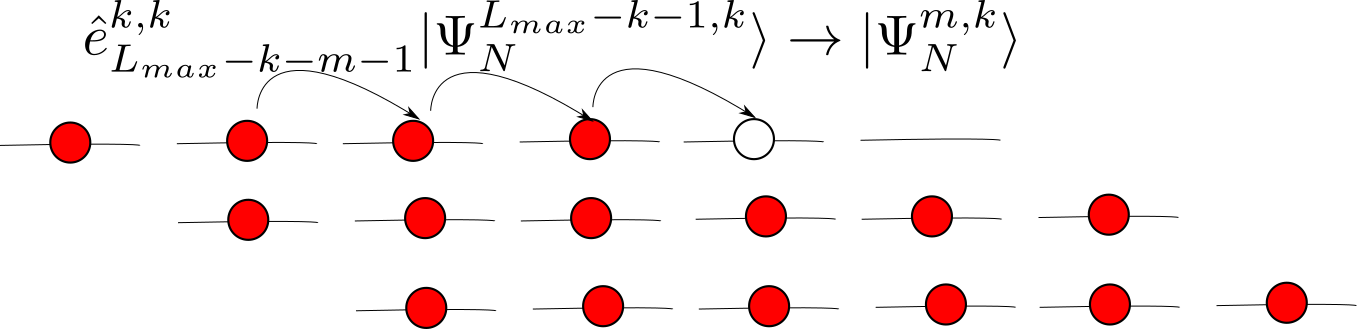}}\\
\subfloat[]
{\includegraphics[width=0.4\textwidth]{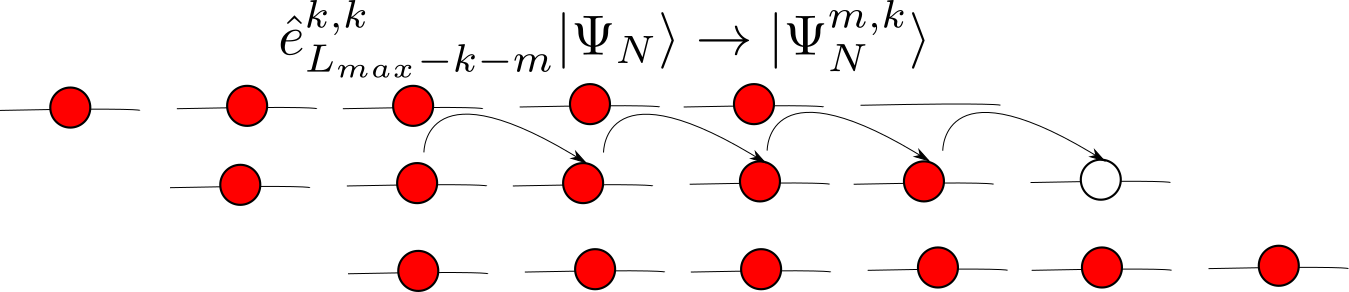}~~~~~~~~~~~}\\
\subfloat[]
{\includegraphics[width=0.4\textwidth]{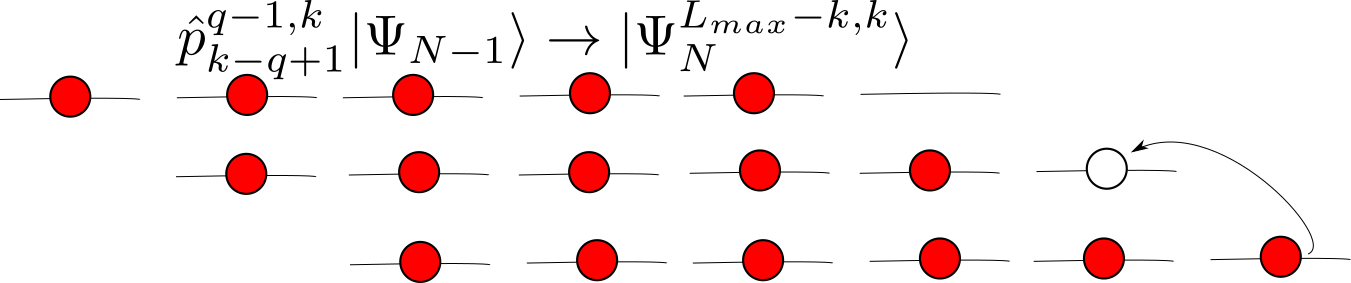}~~~~~~~~~~~}
\subfloat[]{~~~~~~~~~~~\includegraphics[width=0.4\textwidth]{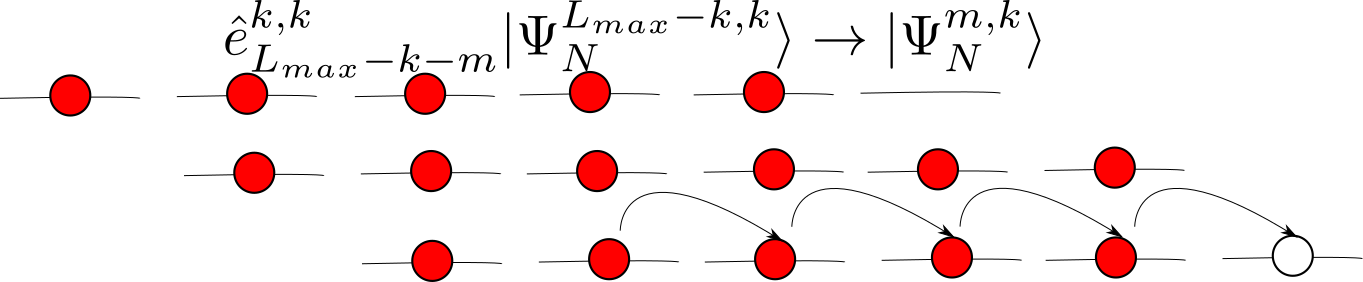}}
\caption{Connecting bare fermion Slater determinants $\ket{\Psi_{{N}-1}}$ (integer quantum Hall) and  $\ket{\Psi^{m,k}_{N}}$ (one hole) via zero mode generators.
Shown are visualizations of the processes used in Eqs.\eqref{tcpsi1}-\eqref{tcpsi2}. All three relevant cases (see main text) are illustrated for $n=3$ Landau levels.}
\label{fig:1}
\end{figure*}

Let us introduce a state $\ket{\Psi^{m,k}_{N}}$,  where we have created a hole in the $ k $th $\Lambda$L at angular momentum $m$ with $ k=0,1...n-1 $.
With Eq. \ref{tcJany}, we have\begin{widetext} \be\tilde{c}_{k,r}\ket{\psi_{N}}=\sum_{m\geq-k}\hat  S_{M(N-1)-r+m}\hat J_{N-1} \tilde c_{k,m}\ket{\Psi_N}=\sum_{m\geq-k}\hat  S_{M(N-1)-r+m}\hat J_{N-1}  \ket{\Psi^{m,k}_{N}}\,.  \ee
Now we need to relate $ \ket{\Psi^{m,k}_{N}} $ to some  zero mode generator acting on $ \ket{\Psi_{N-1}} $, where the only difference between $ \ket{\Psi_{N-1}} $ and $ \ket{\Psi_{N }} $ is that the orbital at $ L_\text{max}-q+1 $ in $ q-1 $th $\Lambda$L in $ \ket{\Psi_{N-1}} $ is vacant.
What's required is that the zero mode generator moves the particle from  the orbital corresponding to $ \tilde c^*_{k,m} $ to that  corresponding to $ \tilde c^*_{q-1,L_{\text{max}}-q+1} $ in $ \ket{\Psi_{N-1}} $. For the sake of conciseness,  we will simply say  moving the particle from $ \tilde c^*_{k,m} $ to $ \tilde c^*_{q-1,L_{\text{max}}-q+1} $ and similarly for other processes involving moves of particles.

As seen in Fig. \ref{fig:1}, we need to consider three cases, (i) $ k>q-1 $, (ii) $ k= q-1 $ and (iii) $ k< q-1 $.
In case (i), $ k>q-1 $, the first step is to act with $ \hat p^{q-1,k}_{k-q+2} $ on  $ \ket{\Psi_{N-1}} $ so that one particle is moved from $ \tilde c^*_{k,L_{\text{max}}-k-1} $ to  $ \tilde c^*_{q-1,L_{\text{max}}-q+1} $. The second step is to further act $ \hat e^{k,k}_{L_{\text{max}}-k-m-1} $ on the resultant state to move all the particles in $ k $th $\Lambda$L beginning with $ \tilde c^*_{k,m} $ and ending with $ \tilde c^*_{k,L_{\text{max}}-k-2} $ to the right such that their angular momenta all increase by 1. This is reflected by the following identity,\be \hat e^{k,k}_{L_{\text{max}}-k-m-1}\hat p^{q-1,k}_{k-q+2}  \ket{\Psi_{N-1}}=(-1)^{f(m)}\ket{\Psi^{m,k}_{N}}, \ee where $ f(m)=(n-q) L_{\text{max}}+m+k$.

This leads to  \be\label{tcpsi1}\tilde{c}_{k,r}\ket{\psi_{N}}=\sum_{m\geq -k} (-1)^{f(m)}\hat  S_{M(N-1)+m-r}\hat e^{k,k}_{L_{\text{max}}-k-m-1}\hat p^{q-1,k}_{k-q+2}  \ket{\psi_{N-1}},\ee where we have used the commutation relations \eqref{Pp} and \eqref{Je}.

Case (ii), $ k= q-1 $,
is very similar, only that no action with a $\hat p$-type operator is necessary.
We obtain  \be\label{tcpsi3}\tilde{c}_{k,r}\ket{\psi_{{N}}}=\sum_{m\geq -k} (-1)^{f(m)}\hat  S_{M({N}-1)+m-r} \hat e^{k,k}_{L_{\text{max}}-k-m}  \ket{\psi_{{N}-1}}.\ee

In case (iii), $ k< q-1 $, the first step is to act with $ \hat p^{q-1,k}_{k-q+1} $ on  $ \ket{\Psi_{N-1}} $ so that one particle is moved from $ \tilde c^*_{k,L_{\text{max}}-k} $ to  $ \tilde c^*_{q-1,L_{\text{max}}-q+1} $. The second step is to further act $ \hat e^{k,k}_{L_{\text{max}}-k-m} $ on the resultant state to move all the particles in $ k $th $\Lambda$L beginning with $ \tilde c^*_{k,m} $ and ending with $ \tilde c^*_{k,L_{\text{max}}-k-1} $ to the right such that their angular momenta all increase by 1.
The overall phase picked up in this process differs by $L_{\text{max}}+1$ from the formulas given for the other two cases.
Thus we have
\be\label{tcpsi2}\tilde{c}_{k,r}\ket{\psi_{{N}}}=\sum_{m\geq -k} (-1)^{f(m)+L_{\text{max}}+1}\hat  S_{M({N}-1)+m-r} \hat e^{k,k}_{L_{\text{max}}-k-m} \hat p^{q-1,k}_{k-q+1} \ket{\psi_{{N}-1}}.\ee

It is now clear that we can repeat the logic that led to the recursion \eqref{Lrec} for the higher $\Lambda$L composite fermion states: to this end, we simply state
\be\label{psi_numberop}
   \ket{\psi_N}=\frac{1}{N}
   \sum_{k,r} c^\dagger_{k,r} c_{k,r} \ket{\psi_N}=\frac{1}{N}
   \sum_{k,r} \tilde c^\ast_{k,r} \tilde c_{k,r} \ket{\psi_N}\,.
\ee
In here, we simply replace
$\tilde c_{k,r} \ket{\psi_N}$ with Eqs.
\eqref{tcpsi1}-\eqref{tcpsi2}.
This gives the desired recursion of $\ket{\psi_N}$ in terms of $\ket{\psi_{N-1}}$. In the following section, we apply these results to the special case $n=2$ again.\end{widetext}

\section{Recursion formulas for  $n=2$ $\Lambda$-level composite fermion states\label{25rec}}
In the last section, we have constructed recursive formulas for any second-quantized composite fermion wave functions. In this section, we will further simplify these formulas for composite fermion states involving two $ \Lambda  $Ls. In Sec. \ref{PZMP}, we will prove,
for the special case $M=2$ describing the Jain $2/5$ state, that this state is indeed the densest zero mode of its parent Hamiltonian of the general form  \eqref{H}.
Our proof differs from a previous one \cite{Chen17} in that it makes no use whatsoever of the polynomial structure of the state's first-quantized wave function, but rests entirely on the operator algebra developed here.
There, we will also  comment further on the connection between the quasihole operators given in Ref. \onlinecite{Chen17} and those in this paper. In a similar vein,  we will show how to extract the filling factor of the CF-states using the present, ``polynomial free'' apparatus. In this section, we will find it convenient to denote the particle number as $2N$ and $2N+1$, respectively, for the even and odd case, as in \Eq{Jain2LL} above.

We now use Eq. \eqref{tcpsi3}, specializing to $n=2  $, $k=1  $, $q=1  $ and $L_\text{max}=N $ for $2N+1 $   particles.
This gives
\begin{widetext}
\be\label{t1si2n1old}\tilde c_{1,r}\ket{\psi_{2N+1}}=  \sum\limits_{m=-1}^{N-2}(-1)^{N+m+1}\hat S_{2MN+m-r}\hat e^{1,1}_{N-2-m}\hat p^{0,1}_{2}\ket{\psi_{2N}}.\ee

The above can be put into a concise form,
\be\label{1si2n1}  c_{1,r}\ket{\psi_{2N+1}}=-\sqrt{(r+1)! }\hat S^{\sharp 1,1}_{(2M+1)N-2-r}\hat p^{0,1}_{2}\ket{\psi_{2N}},\ee
where
\be
\hat S^{\sharp a,b}_{\ell} = \sum\limits_{m} ( - 1)^{m}\hat S_{\ell-m} \hat e^{a,b}_{m}\,,
\ee
a definition that we may adopt for any $n$.
In the last equation,
we have also
replaced $\tilde c_{1,r}$
with the operator $c_{1,r}$, referring to the original (orthonormal) basis.

Similarly using Eq. (\ref{tcpsi2}), we obtain\be\label{t0si2n1}\tilde c_{0,r}\ket{\psi_{2N+1}}=\hat S^{\sharp 0,0}_{(2M+1)N-r}  \ket{\psi_{2N}}.\ee
This leads to \be\label{0si2n1}c_{0,r}\ket{\psi_{2N+1}}=-\sqrt{r!} \left((r+1)\hat S^{\sharp 1,1}_{(2M+1)N-2-r}\hat p^{0,1}_{2}-\hat S^{\sharp 0,0}_{(2M+1)N-r}\right)\ket{\psi_{2N}},\ee Note that 
\begin{equation}\label{lmax}\begin{split} &c_{1,r}\ket{\psi_{2N+1}} = 0 \quad\text{for} \quad r>(2M+1)N-2,\\& c_{0,r}\ket{\psi_{2N+1}}=0 \quad\text{for}\quad r>(2M+1)N,\end{split}\end{equation} as  by definition,  $S^{\sharp a,b}_\ell$ vanishes for $\ell<0$. 
This establishes that the highest occupied orbital in $\ket{\psi_{2N+1}}$ has angular momentum $\ell_{\sf max}\leq (2M+1)N$. 
Moreover, since $S^{\sharp a,b}_0=\mathbb{1}$, \Eq{0si2n1}
for $r=(2M+1)N$ gives that the orbital created by $c_{0,{(2M+1)N}}^\dagger$ is certainly occupied in the state $\ket{\psi_{2N+1}}$, as long as $\ket{\psi_{2N}}$ is not zero. In particular, the state $\ket{\psi_{2N+1}}$ does not vanish as long as $\ket{\psi_{2N}}$ does not. Assuming this for the moment, we find $\ell_{\sf max}=(2M+1)N$. Defining the filling factor as the particle number $2N+1$ divided by $\ell_{\sf max}$, we see that the filling factor approaches $2/(2M+1)$ in the thermodynamic limit, as expected. Similar arguments carry over to larger $n$.

In the same way of obtaining
Eqs. \ref{1si2n1} and \ref{0si2n1}, we obtain

\be\label{1si2n}c_{1,r}\ket{\psi_{2N}}=(-1)^{N+1}\sqrt{ (r+1)! } \hat S^{\sharp 1,1}_{(2M+1)N-M-2-r} \ket{\psi_{2N-1}} \ee and 
\be\label{0si2n}c_{0,r}\ket{\psi_{2N}} =-\sqrt{ r! } \left((-1)^N (r+1)\hat S^{\sharp 1,1}_{(2M+1)N-M-2-r}+ \hat S^{\sharp 0,0}_{(2M+1)N-M-1-r}\hat p^{1,0}_{-1}\right)\ket{\psi_{2N-1}}.\ee
Again, we can immediately see that $c_{1,r}\ket{\psi_{2N}}$ vanishes for $r>(2M+1)N-M-2$ and $c_{0,r}\ket{\psi_{2N}}$ vanishes for $r>(2M+1)N-M-1$. On the other hand,
$c_{0,(2M+1)N-M-1}\ket{\psi_{2N}}$ is proportional to $\hat p^{1,0}_{-1}\ket{\psi_{2N-1}}$. In particular, $\ket{\psi_{2N}}$  is nonzero if $\hat p^{1,0}_{-1}\ket{\psi_{2N-1}}$ is, which follows immediately by acting on the latter with $\tilde c_{1,(2M+1)(N-1)-1}$, commuting past  $\hat p^{1,0}_{-1}$, and using earlier observations for the state of odd particle number. Together with the observation below \Eq{lmax}, this establishes inductively that the states $\ket{\psi_{2N}}$, $\ket{\psi_{2N+1}}$ do not vanish (even if we did not know the meaning of the operator $\hat J_N$ in first-quantization), and that
$\ell_{\sf max}=(2M+1)N$ for $\ket{\psi_{2N+1}}$ and $\ell_{\sf max}=(2M+1)N-M-1$ for $\ket{\psi_{2N}}$.

Now we use Eqs. \ref{1si2n1}, \ref{0si2n1}  and the identity Eq. \ref{psi_numberop} to get a recursive formula
\be
\label{re}
\begin{split}
\ket{\psi_{2N+1}}&=\frac{-1}{2N+1}\sum\limits_{r=-1}^{(2M+1)N-2} \sqrt{(r+1)!}\,c_{1,r}^\dagger \hat S^{\sharp 1,1}_{(2M+1)N-2-r}\hat p^{0,1}_{2}\ket{\psi_{2N}}\\&-\frac{1}{2N+1}\sum\limits_{r=0}^{(2M+1)N-2}(r+1)\sqrt{r!}\,c_{0,r}^\dagger \hat S^{\sharp 1,1}_{(2M+1)N-2-r}\hat p^{0,1}_{2}\ket{\psi_{2N}}\\&+\frac{1}{2N+1}\sum\limits_{r=0}^{(2M+1)N}\sqrt{r!}\,c_{0,r}^\dagger  \hat S^{\sharp 0,0}_{(2M+1)N-r}\ket{\psi_{2N}}. \end{split}
\ee Likewise, we can also obtain  $ \ket{\psi_{2N}} $\ from $ \ket{\psi_{2N-1}} $, \be\label{re2}\begin{split}
\ket{\psi_{2N}}&=\frac{(-1)^{N+1}}{2N}\sum\limits_{r=-1}^{(2M+1)N-M-2}\sqrt{(r+1)!}\,c_{1,r}^\dagger \hat S^{\sharp 1,1}_{(2M+1)N-M-2-r}\ket{\psi_{2N-1}}\\&+\frac{(-1)^{N+1}}{2N}\sum\limits_{r=0}^{(2M+1)N-M-2}(r+1)\sqrt{r!}\,c_{0,r}^\dagger  \hat S^{\sharp 1,1}_{(2M+1)N-M-2-r}\ket{\psi_{2N-1}}\\&-\frac{1}{2N}\sum\limits_{r=0}^{(2M+1)N-M-1}\sqrt{r!}\,c_{0,r}^\dagger \hat S^{\sharp 0,0}_{(2M+1)N-M-1-r}\hat p^{1,0}_{-1}\ket{\psi_{2N-1}}. \end{split}
\ee  \end{widetext} The above recursions, together with the expressions of local charge-1 holes through zero-mode generators acting on an incompressible state, as well as their $n>2$ generalizations of the preceding section, are the central results of this paper.

\section{Proof of zero mode property}\label{PZMP}
The construction of parent Hamiltonians for FQH states has traditionally emphasized analytic clustering properties of special wave functions. Obstructions for successfully doing this, so far, for most composite fermion states have been discussed by some of us \cite{Chen17}.
In short, we argued that a successful parent Hamiltonian satisfying the zero mode paradigm discussed in the introduction
is   possible in principle only for {\em unprojected} CF states, such as discussed in this paper.
(There may, of course, be parent Hamiltonians outside this paradigm \cite{recentJainpaper}.)
On the other hand, Landau level mixing makes it harder to harvest nice analytic clustering properties for the construction of a parent Hamiltonian.
A notable exception is the case $n=M=2$, leading to the Jain 2/5 state. An extensive discussion of its parent Hamiltonian was given in Ref. \onlinecite{Chen17}. There, some of the framework established in this paper has been anticipated, as well as the fact that the zero mode properties of the 2/5-parent Hamiltonian can be understood as a purely algebraic consequence of the second-quantized operators that can be used to define it [\Eq{H2} below] and their interplay with the zero mode generators extensively discussed here. Indeed, this approach allows one to establish properties of parent Hamiltonians while ``forgetting'' the analytic properties of the associated first-quantized many-body wave functions. While this is somewhat counter to traditional construction principles in FQH physics, we argue this to be fruitful in the context of CF states with $n\geq 2$, where parent Hamiltonians are somewhat scarce. This approach also resonates with the manifestly guiding-center-projected language recently advocated by Haldane \cite{Haldane2011}. While in Ref. \onlinecite{Chen17} we did not elaborate on how to establish the zero mode properties of the $2/5$ Hamiltonian in such a purely algebraic manner, here we are in a perfect position to do so. We begin by presenting the Hamiltonian as the sum of four two-particle projection operators at each pair-angular momentum $2R$,

\be\label{H2}\begin{split}
H &= E^{(1)}\sum\limits_R {  \cT_R^{(1)\dag}} {  \cT_R^{(1)}}+ E^{(2)}\sum\limits_R \cT_R^{(2)\dag}   \cT_R^{(2)}\\& + E^{(3)}\sum\limits_R  \cT_R^{(3)\dag}   \cT_R^{(3)}+ E^{(4)}\sum\limits_R   \cT_R^{(4)\dag}   \cT_R^{(4)}.
\end{split}
\ee

Here, $\cT^{(\lambda)}_R=\sum_{x,m_1,m_2} \eta^{(\lambda)}_{R,x,m_1,m_2}c_{m_1,R-x}c_{m_2,R+x} $ is a fermion bilinear that destroys a pair of particles of angular momentum $2R$. The details of the form factors  $\eta^{(\lambda)}_{R,x,m_1,m_2}$ are of no importance in the following, but will be given in Appendix \ref{z}.
The $E^{(\lambda)}$ are positive constants that are arbitrary in principle, but may be chosen so as to give the Hamiltonian a simple ``Trugman-Kivelson'' form \cite{TK} in first-quantization, see again Appendix \ref{z} for this choice. Note that the sum over $R$ goes over integers and half-odd integers, and $x$ sums in the $\cT$ operators are restricted so that $R\pm x$ are integers.

From the positivity of each of the four terms in the Hamiltonian \eqref{H2}, it follows that the zero mode property is equivalent to the following:
\be\label{zero}\cT_R^{(\lambda)}\ket{\psi_{\text{zm}}} =0,\quad \mbox{for} \;\; \lambda=1,2,3,4. \ee

The zero mode property of the  Jain $2/5$ state as given by \Eq{genCFstate} (for $M=n=2$), with the {\em recursively defined} composite fermion operator $\hat J_N$, \Eq{Jgen}, then rests on the following properties.

(1). The operators identified in Sec. \ref{ZMGs} are zero mode generators precisely in the strict sense defined at the end of Sec. \ref{OR}: Namely, they leave invariant the zero mode space defined in terms of the Hamiltonian through \Eq{zero}. We show this in Appendix \ref{z}.

(2). The operators $\cT^{(\lambda)}_R$ satisfy
\be\label{prop2}
   \cT_R^{(\lambda)}=\frac 12 \sum_{m,k} [\cT_R^{(\lambda)},c_{m,k}^\dagger]c_{m,k}\,.
\ee
This is a generic property of the fermion bilinears, and does not depend on the form factors $\eta^{(\lambda)}_{R,x,m_1,m_2}$.

(3). The two-particle CF state $\ket{\psi_{N=2}}$ is a zero mode, allowing an ``induction beginning''.

We begin by demonstrating property 3. Since $\ket{\psi_{N=0}}=\ket{0} $, we get $\ket{\psi_{N=1}}=c_{0,0}^\dagger \ket{0} $ and $\ket{\psi_{N=2}}=(\sqrt{2}c_{1,-1}^\dagger c_{0,2}^\dagger +2c_{0,1}^\dagger c_{1,0}^\dagger -\sqrt{2}c_{0,0}^\dagger c_{1,1}^\dagger -4c_{0,0}^\dagger c_{0,1}^\dagger)\ket{0} $ using Eqs. (\ref{re}) and (\ref{re2}).
It is trivial to see that $\ket{\psi_{N=0}}  $, and $\ket{\psi_{N=1}}  $ are zero modes, and indeed $\ket{\psi_{N=2}}$ can also be straightforwardly shown to satisfy the zero mode conditions Eq. (\ref{zero}), using the explicit formulas for the $\cT^{(\lambda)}_R$ given in Appendix \ref{z}. (Note that this only requires the relatively simple special cases with $R=1/2$.)   Now assuming
$\ket{\psi_{2N}} (N\geq1) $ is a zero mode, we  immediately find \begin{subequations}
\label{induct}
\be T_R^{(\lambda)}c_{1,k}\ket{\psi_{2N+1}}=0\ee and \be T_R^{(\lambda)}c_{0,k}\ket{\psi_{2N+1}}=0,\ee
\end{subequations}
since on the right-hand sides of Eqs. \eqref{1si2n1} and \eqref{0si2n1}, all operators are zero mode generators, acting on the zero mode $\ket{\psi_{2N}}$, thus giving another zero mode.

Acting  with $T_R^{(\lambda)}$,  $ \lambda=1,2,3,4 $ on the identity  \eqref{psi_numberop} with particle number being $2N+1$ instead of $N$, and then using \Eq{induct}, we obtain
\be\begin{split}\cT_R^{(\lambda)}\ket{\psi_{2N+1}}&=\frac{1}{2N+1}\sum_k [\cT_R^{(\lambda)},c_{0,k}^\dagger]c_{0,k}\ket{\psi_{2N+1}}\\&+\frac{1}{2N+1}\sum_k [\cT_R^{(\lambda)},c_{1,k}^\dagger]c_{1,k}\ket{\psi_{2N+1}}\\&=\frac{2}{2N+1}\cT_R^{(\lambda)}\ket{\psi_{2N+1}},
\end{split}\ee
where in the last line, we have used \Eq{prop2}.
This implies that $ \ket{\psi_{2N+1}} $ satisfies the zero mode condition Eq. \eqref{zero}.
The induction step from odd particle number $2N+1$ to even particle number $2N+2$ proceeds analogously, with the help of Eqs. \eqref{1si2n} and \eqref{0si2n}, thus concluding the induction proof for the zero mode property of $n=M=2$ ($\nu=2/5$) Jain state. Using the methods of Ref. \onlinecite{Chen17}, which we later
characterized  as making use of an ``entangled Pauli principle''(EPP) \cite{sumanta18}, we can also establish that these are the densest possible (highest filling factor) zero modes (see Ref. \onlinecite{sumanta18} for details). Aside from the EPP, the only ingredients needed are knowledge of the total angular momentum of the CF state as defined in \Eq{genCFstate}, and/or its highest occupied orbital, all of which is either manifest or follows from the discussion in Sec. \ref{25rec}. In particular, as we have shown here, none of this requires knowledge of the analytic structure of the first-quantized Jain 2/5 state wave function.

One may envision that the results of this section readily generalize to other CF states, for which, to the best of our knowledge, so far no (zero mode paradigm) parent Hamiltonians have been discussed in the literature, with the exception of the case $n=1$. This requires identification of the proper set of operators $\cT^{(\lambda)}$ that generalize the algebraic features discussed here and in Appendix \ref{z} to larger $n$ and $M$, which will require a larger set of such operators. We will comment on this interesting problem elsewhere \cite{SumantaToBePublished}.

\section{Microscopic Bosonization\label{mbos}}

In this brief section, we make contact with an observation made in Ref. \onlinecite{Chen17} (and earlier for Laughlin states in Ref. \onlinecite{Mazaheri14}). This is the fact that the zero mode generators $\hat p_k^{m,m}$ (no summation implied), \Eq{Pab}, formally look like bosonic modes generating excitations    in the $m$th branch of a free chiral fermion edge theory. Indeed, a zero mode at small angular momentum $k$ relative to the incompressible ground state must be interpreted as a low-energy edge excitation.
This can be made concrete by considering a confining potential proportional to total angular momentum, which may be added to the parent Hamiltonian --- in those cases where one is known --- without changing the eigenstates of the system. Our result can then be considered a microscopic form of bosonization --- the identification of generators of eigenstates for the {\em microscopic} Hamiltonian with corresponding counterparts in the effective edge theory. To make this case, we must argue that the $\hat p_k^{m,m}$ in some sense generate a {\em complete set} of low-energy modes. In this case, we can unambiguously deduce the effective edge theory from exact properties of the microscopic parent Hamiltonian. We note that the latter is quite non-trivial even for the Laughlin state parent Hamiltonians using conventional polynomial methods \cite{Stone90}.

In Ref. \onlinecite{Chen17}, we conjectured that the operators formed by products of the $\hat p_k^{a,b}$ do indeed generate a complete set of zero modes (not just at small angular momentum) for the Jain $2/5$ parent Hamiltonian when acting on the Jain $2/5$ state
$\ket{\psi_N}$. With the results of this work, this becomes an easy corollary.
To this end, we first note that a complete set of zero modes is given by
\be\label{generic}
  \hat J_N\ket{\Phi},
\ee
where $\ket{\Phi}$ is any $N$-particle state within the first $n$ LLs. Specifically for the Jain $2/5$ state ($n=M=2$), we established the densest zero mode in the preceding section, which is of the form \eqref{generic}. The general statement for {\em all possible} zero modes can either be established in first-quantization, or, using EPP-based methods and knowledge of the densest zero mode, in second-quantization. See Ref. \onlinecite{Chen17} for details. Here we want to show that all zero modes, of given total particle number $N$, are obtained by acting on the densest zero mode, $\ket{\psi_N}=\hat J_N \ket{\Psi_N}$,
\Eq{genCFstate}, with sums of products of the operators $\hat p_k^{a,b}$. (For $n=1$, pertinent considerations were carried out earlier \cite{Mazaheri14}, using somewhat different methods.) We first focus on such zero modes where the $\ket{\Phi}$ in \Eq{generic} has the same particle number {\em in each $\Lambda$ level} as the integer quantum Hall state $\ket{\Psi_N}$. For this we may restrict ourselves to the operators $\hat p^{a,b}_k$. Since we have established that these operators commute with $\hat J_N$, the statement is thus simply that each fermion state $\ket{\Phi}$, with given particle number in each of $n$ $\Lambda$Ls equal to that in the state $\ket{\Psi_N}$, can be expressed as  $\ket{\Psi_N}$ acted upon by sums of products of the $\hat p^{a,b}_k$. For $a=b=m$, these operators now act on $\Lambda$Ls {\em exactly} as the ones that appear in the bosonization dictionary.
The fact that these operators, within each branch ($\Lambda$L) $m$, generate the full fermionic subspace of the same particle number when acting on the ``vacuum'' present in $\ket{\Psi_N}$ is a well-known theorem in bosonization. Here, we need a version of this theorem at finite particle number, which is also readily available \cite{schonhammer, Mazaheri14}.
Similarly, it is easy to see that the operators $\hat p^{a,b}_k$, $k\geq b-a$, which likewise commute with $\hat J_N$, can be used to generate an arbitrary imbalance in particle number between the occupied $\Lambda$Ls in $\ket{\Psi_N}$, without introducing any holes into any of these $\Lambda$Ls. By the same reasoning, when acting on these states with all possible combinations of the $\hat p^{m,m}_k$, we generate the full Fock space of $n$ $\Lambda$Ls at fixed particle number. Note that the relative ease with which we can establish this property here crucially depends on having control of the relationship between the operator $\hat J_N$ and the operators $\hat p^{a,b}_k$, in particular their trivial commutators.

The above considerations may serve as an alternative proof \cite{Chen17}  for the fact that the Jain $2/5$ parent Hamiltonian falls into the ``zero mode paradigm'': Counting of zero modes at given angular momentum $\Delta k$ relative to the ``incompressible state'' (densest zero mode) reproduces exactly the mode counting in an associated conformal edge theory.

\section{Composite fermion state order parameters\label{OPsec}}

The question of off-diagonal long-range order has been an influential subject in the theory of the Hall effect, leading, in particular, to a description in terms of effective Ginzburg-Landau type actions \cite{GirvinMacDonald, HROP, readOP, ReadOP2}. Beyond this theoretical use, non-local order parameters could in principle be useful in practical numerical calculations, serving as diagnostics for the myriad possible phases in the fractional quantum Hall regime. Unfortunately, a number of reasons seem to have prohibited widespread use of this approach.
For one, there is the problem of efficient evaluation of non-local objects such as
\be\label{OP}
  {\cal O}(z):= (\Psi(z)^\dagger)^p\prod_i(z-z_i)^q\,,
\ee
where the $z_i$ are the complex electron coordinates, and $\Psi(z)^\dagger$ is a local electron creation operator. This order parameter is expected to characterize the order of all composite fermion states with ``single-particle condensates'' at filling fraction $\nu=p/q$ \cite{readOP, ReadOP2}.  
The non-locality of this object and the mixed first/second-quantized definition make numerical evaluation challenging, though, making use of special properties of spherical geometry, related order parameters have been evaluated for eight-particle systems \cite{HROP}. We are not aware of any attempt to numerically evaluate \Eq{OP} on the cylinder, which is arguable the preferred geometry for DMRG. What is more important, the order parameter \eqref{OP} is by itself still a rather crude diagnostic. Already for composite fermion states in $n$ $\Lambda$Ls, a multiplet of $n$ independent order parameters is expected to exist, which can be given precise meanings in suitable variational wave functions \cite{ReadOP2}, and which are the basis for field theoretic and/or Ginzburg-Landau level descriptions \cite{ReadOP2, WenBlok90}. Except for \Eq{OP}, which is always a member of the ``lattice'' \cite{ReadOP2} of order parameters, we are, however, not aware of a general definition of these order parameters as operators acting on the microscopic Fock space.

The results of the preceding sections allow us to address these obstacles in the following way. We will be able to express order parameters such as \Eq{OP} in a fully second-quantized form that is directly applicable to planar, spherical, and cylinder geometries, respectively. What's more, for $n>1$ composite fermion states we will do the same for an $n$-tuplet of generators of the order parameter lattice, all of whose members will create charge 1 and are thus more elementary than \Eq{OP}, which creates charge $p>1$ for $n>1$.

A close connection between quantum Hall-type order parameters and the developments of this paper could be surmised on the basis that Read wrote the Laughlin state as $(\int dz ({\cal O}(z))^N|{\sf vac}\rangle$, which leads to the Laughlin state recursion \Eq{Lrec}, albeit in a mixed first/second-quantized guise. We will immediately discuss the general case $n\geq 1$. We start with an argument similar to one made by Read \cite{readOP} for the Laughlin state and, originally, leading up to the special order parameter \eqref{OP}.
We will, however, start by working in the orbital basis. Consider the correlation function of the orbital density $\rho_r= \sum_k c_{k,r}^\dagger c_{k,r}$,
\be\label{order1}
\langle \psi_{N+1} |\rho_r \rho_{r'}|\psi_{N+1}\rangle
\longrightarrow \langle\rho_r\rangle
\langle\rho_{r'}\rangle\sim \nu^2
\,,
\ee
where, on the right-hand side, we take the limit of large $|r-r'|$ and expect that correlations decay exponentially, causing the un-connected correlator to approach a non-zero constant equal to the square of the filling factor $\nu$.
As argued by Read,  electron destruction operators such as $c_{k,r}$ acting on $\ket{\psi_{N+1}}$ generally should give a state that can be thought of as $q$ quasi-hole operators, fused at the same location, acting on the incompressible state $\ket{\psi_N}$. Here we use the fact that in the presence of a special Hamiltonian as discussed above, this notion becomes entirely sharply defined in a microscopic sense.
Indeed, since $\ket{\psi_N}$ is a zero mode of the Hamiltonian, then so is $c_{k,r}\ket{\psi_{N+1}}$, as all the $c_{k,r}$ commute with all fermion bilinears $\cT_R^{(\lambda)}$.
 $c_{k,r}\ket{\psi_{N+1}}$ is thus always {\em uniquely} expressible in any basis of $N$-particle zero modes. Moreover, one may prefer to think of $N$-particle zero modes as being generated by appropriate operators acting on the $N$-particle incompressible state. While in some abstract sense, such operators may always exist, here we have already unambiguously defined them via concrete expressions involving only microscopic electron creation and annihilation operators, Eqs. \eqref{tcpsi1}-\eqref{tcpsi2}. More concisely, we have shown that \be
  \tilde c_{k,r}\ket{\psi_{N+1}}  =
  S^{\sharp k,k}_{MN-r-\delta+L_{{\sf max}}(N+1,n) }R_{N,n,k} \ket{\psi_N},
\ee
where $R_{N,n,k}$ is a local operator (in the orbital basis) that may be inferred from Eqs. \eqref{tcpsi1}-\eqref{tcpsi2}, along with $\delta\in\{0,1\}$.
Writing $\rho_r= \sum_k \tilde c_{k,r}^\ast\tilde c_{k,r}$ as in \Eq{psi_numberop}, \Eq{order1} takes on the form
\be\label{order2}
\langle \psi_{N} |{\cal O}_{r}^\dagger {\cal O}_{r'}|\psi_{N}\rangle
\longrightarrow \nu^2
\,,
\ee
where
\be\label{OP2}
{\cal O}_{r}=\sum_k    \tilde c_{k,r}^\ast  S^{\sharp k,k}_{MN-r-\delta+L_{{\sf max}}(N+1,n) }R_{N,n,k}\,.
\ee
The object ${\cal O}_{r}$ therefore exhibits off-diagonal long-range order (ODLRO).
Equation (\ref{OP2}) is closely related to \Eq{OP} only for
 $p=1$.
 It is different for $p>1$, as it adds only one particle overall whereas \Eq{OP} adds $p$ particles.
More importantly, \Eq{OP} is just a single point in an ``order parameter lattice'' that has $n$ generators \cite{ReadOP2}.
In contrast, it stands to reason that in \Eq{OP2} each term for given $k$ contributes to the ODLRO. In fact,  this is of a kind with an SU($n$) symmetry discussed in Ref. \onlinecite{ReadOP2} on the basis of variational wave functions, and which moreover can be seen to be a property of the zero mode spaces associated to {\em all} composite fermion states, given appropriate parent Hamiltonians \cite{SumantaToBePublished}. (This is quite a robust property of $n>1$ special Hamiltonians, and generalizes even to more complicated ``parton'' states \cite{sumanta18}.) It is thus natural to define
\be\label{OP3}
{\cal O}_{k,r}=    \tilde c_{k,r}^\ast  S^{\sharp k,k}_{MN-r-\delta+L_{{\sf max}}(N+1,n) }R_{N,n,k}\,.
\ee
and identify this family of $n$ operators for $k=0\dotsc n-1$ as the generators of the order parameter lattice, which exhibit ODLRO  in the orbital degree of freedom $r$. In fact, we can make this argument more directly by noting that the ORDLO of
\Eq{OP3}, for the composite fermion states $\ket{\psi_N}$, follows exactly in the same manner as for the original ${\cal O}_r$, assuming only that the ``partial densities'' $\rho_{k,r}= \tilde c^\ast_{k,r} \tilde c_{k,r}$ have exponentially decaying correlations (which is  given \cite{hastingspaper} in the presence of a gap), and assume a non-zero expectation value $\langle\rho_{k,r}\rangle\neq 0$.
Note that we have $\rho_r= \sum_{k=0}^{n-1} \rho_{k,r}$, and the $\rho_{k,r}$ essentially measure the occupancy density in the $k$th $\Lambda$-level as defined above.

Several remarks are in order. For one, the operators $R_{N,n,k}$ are a consequence of choosing a particular edge configuration for the reference state $\ket{\psi_N}$, which is not uniquely determined in general without some conventions, such as chosen above. These operators must be kept if \Eq{order2} is to be {\em exact} for the given composite fermion state $\ket{\psi_N}$ as defined above. However, the ODLRO is expected to be a property of all states in the same phase, and is not expected to rely on the choices leading to the $R_{N,n,k}$-operators. (Note that in \Eq{tcpsi3}, $R_{N,n,k}$ is proportional to the identity anyway, and is proportional to the single body operators $\hat p^{q-1,k}_{k-q+2}$, $\hat p^{q-1,k}_{k-q+1}$ in the other cases, respectively.)  In the same vein, the parameter $\delta\in\{0,1\}$ is irrelevant to the ODLRO. We may thus settle for the slightly more streamlined variant
\be\label{OP4}
{\cal O}'_{k,r}=    \tilde c_{k,r}^\ast  S^{\sharp k,k}_{MN-r+L_{{\sf max}}(N+1,n) }\,.
\ee
Note that although we have arrived at a reasonably compact definition for these operators using an $n$-Landau level framework, all of these operators remain meaningful, non-trivial, and independent when projected onto the lowest Landau level. To see this, observe that the $S^\sharp$ operator in \Eq{OP4} creates a (charge 1) quasi-hole in the $k$th composite fermion $\Lambda$-Level, at orbital location $r$. One expects such states for different $k$ to remain linearly independent  even after lowest-LL projection.  For an $n>1$ composite fermion state there are $n$-distinct ways of creating a charge $1$ hole at given (orbital or real space) location.
These $n$ distinct way are encoded in the $S^\sharp$-operators, whose relation to electron creation/annihilation operators is explicitly given here, and which remain distinct objects whether or not we choose to lowest-Landau-level-project.
In the spirit of Ref. \onlinecite{ReadOP2}, to create an order parameter, these $n$ distinct types of holes can then be filled by the action of any electron creation operator, in particular, one in the lowest Landau level.
Note that in particular the creation operator $\tilde c_{k,r}^\ast$ of \Eq{OP4} always has a non-zero component in the lowest Landau level. The relevance of the order parameters \eqref{OP4} is thus by no means limited to the mixed-Landau-level setting used here to derive them. After lowest-Landau-level-projection, the $k$-labels refer to $\Lambda$-levels in the original, purely emergent sense of the term \cite{jainbook}.

It should be emphasized that the two processes in \eqref{OP4} are very different, where the $S^\sharp$-operator creates a hole via flux insertion into one of the $\Lambda$-levels, but without changing overall particle number, a highly non-local operation. In contrast, this hole is then filled by a local electron creation operator. In \Eq{OP4}, both the hole and the subsequently inserted particle are localized in orbital space. If desired, it is easy to construct  corresponding order parameters with both electron and hole localized in real space (but the latter still facilitated by a non-local operator). If a local electron destruction operator $\hat \psi_j(z)$ is obtained via
\begin{equation}
    \hat \psi_j(z)=\sum_{k,r} \,{\cal F}_{k,r,j}(z) \tilde c_{k,r}\,,
\end{equation}
where $ {\cal F}_{k,r,j}(z)$ depends in straightforward ways on the matrix $A(r)_{ab}$ defined in \Eq{tsum} and the Landau level basis wave functions, the desired order parameter is given by
\begin{equation}\label{OP5}
{\cal O}'_j(z)= \hat\psi_j^\dagger(z) \sum_{k,r} {\cal F}_{k,r,j}(z)S^{\sharp k,k}_{MN-r+L_{{\sf max}}(N+1,n) }\,.
\end{equation}
Equation (\ref{OP5}) is obtained following strictly the same logic leading up to \Eq{OP4} \footnote{We could, of course, have obtained an analogous expression based on \Eq{OP3}.}. However, since for $j>0$, $\hat\psi_j^\dagger(z)$ now does create a state orthogonal to the lowest Landau level, projection to the lowest Landau level now warrants replacement of $\hat\psi_j^\dagger(z)$ with $\hat\psi_0^\dagger(z)$. In view of the discussion above, this should not affect the ODLRO of these operators.

\section{Conclusion\label{conclusion}}

In this work, we have developed a comprehensive formalism to discuss composite fermions in Hilbert space. The heart of this formalism is a presentation of the Laughlin-Jastrow flux attachment operator in terms of second-quantized electron creation and annihilation operators. This allows us in particular to define certain operations that {\em add} fermions to a reference composite fermion ground state, as well as general operations that remove them, while staying in the composite fermion sector of the Hilbert space. As a result, we can define Jain composite fermions states recursively in the orbital basis, generalizing similar recursions for Laughlin states. This operator-based approach has several advantages. The properties of parent Hamiltonians, where they exist, can be rigorously established. This in particular establishes edge theories microscopically on much more than variational grounds. $n$-component order parameters for the Jain composite fermion phases can be microscopically defined, i.e., their relation to microscopic electron creation and annihilation operators is fully specified, and their meaning thus extended from a variational subspace to the full Hilbert space.

We expect that this work will spur further developments in particular along several interesting directions: One is the construction of new special parent Hamiltonians for mixed Landau-level wave functions. This includes all of the Jain states \cite{SumantaToBePublished}, 
but also other, more exotic quantum Hall states including parton states \cite{Jain2, wenparton, Jain221, sumanta18}.

Indeed, the present work and the treatment \cite{sumanta18} by some of us of the non-Abelian Jain 221 state can both be   regarded as different natural extensions of earlier work on the Jain $2/5$ state \cite{Chen2014}. It therefore   seems likely that further extensions of the formalism developed here to non-Abelian states are possible. This formalism, in connection with the idea of ``entangled Pauli principles'' (EPP) that naturally extends the notion of ``generalized Pauli principles''  \cite{BH1,BH2} or thin torus patterns \cite{BK,BK1,BK2,BK3,seidel05, seidel2006abelian, SY1, Seidel10,  SY2, Flavin12, zhou2013heat}, represent a powerful new framework to construct and study FQH parent Hamiltonians from the point of view of infinite-range  frustration-free one-dimensional lattice models, as opposed to analytic wave functions. This may further turn out to be beneficial when studying spectral properties of such models at non-zero energy \cite{Amila15}, 
or making connection between EPPs and braiding statistics \cite{seidel2007domain, seidel2008pfaffian,FlavinPRX}.
Another exciting prospect
is the further development of non-local order parameters as numerical diagnostic and theoretical tool. There is further much to be said about the connection between the present developments and the conformal field theory \cite{MR, Hansson_RMP_2017}/matrix-product-state \cite{dubail, zaletel, Estienne} representability of fractional quantum Hall states. We leave these as interesting problems for future work \cite{Schossler_Bandyopadhyay_Seidel_Tobepublished}.

\appendix
\section{Proof of Eq. \ref{cJ}}\label{proof}We first prove Eq. \ref{cJ} by induction. It is trivial to see that it is satisfied for $ N=0,1 $. Now assume \be\label{cJ-} c_r \hat  J_{N-1}=\sum\limits_{m}\hat  S_{M(N-2)-r+m}\hat J_{N-2} c_{m }  \ee is true. The induction hinges on the following two identities,
\be\label{cs} c_r \hat  S_\ell=\sum\limits_{k=0}^{M} (-1)^k {M \choose k}\hat  S_{\ell-k}c_{r-k},\ee \be\label{Sc+} \hat  S_\ell c_r^\dag=\sum\limits_{k=0}^{M} (-1)^k {M \choose k}c_{r+k}^\dag \hat  S_{\ell-k},\ee
which one easily obtains from the definition of the $\hat  S_\ell$ operators, \Eq{Sell}, with the aid of the following two commutators,
\be [ c_r, \hat  e_n]=\hat  e_{n-1}c_{r-1},\ee \be[ \hat  e_n, c_r^\dag]= c_{r+1}^\dag \hat  e_{n-1}.\ee

Then, using the definition in Eq. \ref{JLL} and the identity Eq. \ref{cs} we have \be \begin{split}
c_r \hat  J_N&=\frac{1}{N}\sum\limits_{m}\hat  S_{M(N-1)-r+m}\hat J_{N-1}c_m\\&-\frac{1}{N}\sum\limits_{m,r'}\sum\limits_{k=0}^{M}(-1)^k {M \choose k}c_{r'+m}^\dag \hat  S_{M(N-1)-r'-k}\\&\times c_{r-k}\hat J_{N-1}c_m.\end{split}\ee Henceforth, the indices of sums $ r,r',m,m' $ go from $ 0 $ to $ +\infty $ unless otherwise noted. We can separate the above sum in $ k $ from 0 to $M$ into two partial sums(one is from 0 to $M-1$ and another is   $k=M$) and then use Eq. \ref{cJ-} to get \be\begin{split}
c_r \hat  J_N&=\frac{1}{N}\sum\limits_{m}\hat  S_{M(N-1)-r+m}\hat J_{N-1}c_m\\&-\frac{1}{N}\sum\limits_{m',m,r'}\sum\limits_{k=0}^{M-1}(-1)^k {M \choose k}c_{r'+m}^\dag \hat  S_{M(N-1)-r'-k}\\&\times \hat  S_{M(N-2)-r+k+m'} \hat J_{N-2}c_{m'} c_m\\&-\frac{1}{N}\sum\limits_{m',m,r'} c_{r'+m}^\dag \hat  S_{M(N-1)-r+m'} \hat  S_{M(N-2)-r'}\\&\times \hat J_{N-2}c_{m'}c_m.
\end{split} \ee  In the third term of the above, we have exchanged the order of two commuting $\hat  S $ operators. We can further move $\hat  S_{M(N-1)-r+m'} $  to the left of  $c_{r'+m}^\dag$ using the identity Eq. \ref{Sc+}. After doing this, we have  \be\begin{split}
c_r \hat  J_N &=\frac{1}{N}\sum\limits_{m}\hat  S_{M(N-1)-r+m}\hat J_{N-1}c_m\\& -\frac{1}{N}\sum\limits_{m',m,r'}\sum\limits_{k=0}^{M-1}(-1)^k {M \choose k} c_{r'+m}^\dag \hat  S_{M(N-1)-r'-k}\\&\times \hat  S_{M(N-2)-r+k+m'}\hat J_{N-2}c_{m'} c_m\\&+\frac{1}{N}\sum\limits_{m'}\hat  S_{M(N-1)-r+m'}\\&\times\Big(\sum\limits_{m,r'} c_{r'+m}^\dag  \hat  S_{M(N-2)-r'}\hat J_{N-2}c_m\Big) c_{m'}\\&+\frac{1}{N}\sum\limits_{m',m,r'}\sum\limits_{k=1}^{M}(-1)^k {M\choose k}c_{r'+m+k}^\dag \hat  S_{M(N-2)-r'}\\& \times \hat  S_{M(N-1)-r+m'-k}\hat J_{N-2}c_{m'} c_m.\end{split}\ee
The third term in the above is just  \be \frac{N-1}{N}\sum\limits_{m'}\hat  S_{M(N-1)-r+m'}\hat J_{N-1}c_{m'}\ee using  Eq. \ref{JLL}. Combined with the first term, it gives the desired result. The second term cancels with the fourth term after we make the change of variables $ k=M-k',r'=r''-k=r''-M+k'$ in  the fourth term and use the fact that $ \hat  S_\ell  \equiv 0 $ for $l>(N-2)M  $ when acting on states with particle number $N-2$. This concludes our induction proof of Eq. \ref{cJ}.

Furthermore, generalizing  the above  proof of Eq. \ref{cJ} to the case of $n$   Landau levels by using notations in Eq. \ref{tsum} with $A(r)$  given in Appendix \ref{AppA} and using the following generalization of Eqs. \ref{cs} and \ref{Sc+},
\be \tilde c_{a,r} \hat  S_\ell=\sum\limits_{k=0}^{M} (-1)^k {M \choose k}\hat  S_{\ell-k}\tilde c_{a,r-k},\ee \be\hat  S_\ell \tilde c^*_{a,r}=\sum\limits_{k=0}^{M} (-1)^k {M \choose k}\tilde c^*_{a,r+k} \hat  S_{\ell-k},\ee we easily arrive at Eq. \ref{tcJany} using the same method.

\section{Zero Mode Generators}\label{z}
In Ref. \onlinecite{Chen17}, we have obtained in second-quantized form the parent Hamiltonian for the unprojected Jain 2/5 state, \be\begin{split}
H &= E^{(1)}\sum\limits_R {  \cT_R^{(1)\dag}} {  \cT_R^{(1)}}+ E^{(2)}\sum\limits_R \cT_R^{(2)\dag}   \cT_R^{(2)}\\& + E^{(3)}\sum\limits_R  \cT_R^{(3)\dag}   \cT_R^{(3)}+ E^{(4)}\sum\limits_R   \cT_R^{(4)\dag}   \cT_R^{(4)},
\end{split}
\ee
where $E^{(1)}=\frac{5+\sqrt{17}}{16\pi}$, $E^{(2)}=\frac{9}{8\pi}$,  $E^{(3)}=\frac{1}{4\pi}$,  $E^{(4)}=\frac{5-\sqrt{17}}{16\pi}$.

\begin{widetext}The bilinear $\cT$-operators are given by $\cT^{(\lambda)}_R=\sum_{x,m_1,m_2} \eta^{(\lambda)}_{R,x,m_1,m_2}c_{m_1,R-x}c_{m_2,R+x} $ with \be\begin{split} \eta^{(1)}_{R,x,m_1,m_2}=&\frac{\sqrt{2}}{2\sqrt {17-\sqrt{17}}}\left( \frac{(-1+\sqrt{17})}{2^{R + 1/2}} \sqrt {  2R + 1 \choose R + x } \, \delta_{m_1,1}\delta_{m_2,0}-\frac{4x}{2^{R + 1/2}} \sqrt {\frac{1}{2R+2}  {2R +2 \choose R +1+ x}  }\, \delta_{m_1,1}\delta_{m_2,1} \right),\\
\eta^{(2)}_{R,x,m_1,m_2}=&\frac{1}{ 2^R 3}\Big( \sqrt{2}\,x  \sqrt {\frac{1}{R}  {2R \choose R + x } } \,\delta_{m_1,0}\delta_{m_2,0} +2 (2x^2-2x-R) \sqrt{\frac{1}{2R(2R+1)}  {2R+1 \choose R + x } } \,\delta_{m_1,1}\delta_{m_2,0}\\&-(2x^3-(3R+2)x) \sqrt {\frac{1}{2R(2R+1)(2R+2)}  {2R+2 \choose R +1+ x } }\,\delta_{m_1,1}\delta_{m_2,1}\Big),\\
\eta^{(3)}_{R,x,m_1,m_2}=&\frac{1-2x}{2^{R + 1/2}} \sqrt{\frac{1}{2R+1}  {2R +1 \choose R + x } } \,\delta_{m_1,1}\delta_{m_2,0},\\
\eta^{(4)}_{R,x,m_1,m_2}=&\frac{\sqrt{2}}{2\sqrt {17+\sqrt{17}}}\left( \frac{(-1-\sqrt{17})}{2^{R + 1/2}} \sqrt {  2R + 1 \choose R + x }  \,\delta_{m_1,1}\delta_{m_2,0}-\frac{4x}{2^{R + 1/2}} \sqrt {\frac{1}{2R+2}  {2R +2 \choose R +1+ x}  } \,\delta_{m_1,1}\delta_{m_2,1} \right).\end{split}\ee
We have found four classes of one-body zero mode generators in Ref. \onlinecite{Chen17}, which leave invariant the zero mode space of the above Hamiltonian,
\be\begin{split} \hat P_d^{(1)}&=\sum_{r=-1}^{+\infty}\sqrt{\frac{(r+d)!}{(r+1)!}}c^\dagger_{0,r+d}c_{1,r},\\ \hat P_d^{(2)}&= \sum_{r=0}^{+\infty} \sqrt{\frac{(r+d)!}{r!}}c^\dagger_{0,r+d}c_{0,r} +\sum_{r=-1}^{+\infty}\sqrt{\frac{(r+d+1)!}{(r+1)!}}c^\dagger_{1,r+d}c_{1,r},\\ \hat P_d^{(3)}&=\sum_{r=-1}^{+\infty} \Big((r+d+1)\sqrt{\frac{(r+d)!}{(r+1)!}}c^\dagger_{0,r+d}c_{1,r} +\sqrt{\frac{(r+d+1)!}{(r+1)!}}c^\dagger_{1,r+d}c_{1,r}\Big),\\\hat P_d^{(4)}&=\sum_{r=0}^{+\infty} \Big(\sqrt{\frac{(r+d+1)!}{r!}}c^\dagger_{1,r+d}c_{0,r}+(r+d+1)\sqrt{\frac{(r+d)!}{r!}}c^\dagger_{0,r+d}c_{0,r}\Big)\\ &-\sum_{r=-1}^{+\infty} \Big((r+1)\sqrt{\frac{(r+d+1)!}{(r+1)!}}c^\dagger_{1,r+d}c_{1,r}+(r+1)(r+d+1)\sqrt{\frac{(r+d)!}{(r+1)!}}c^\dagger_{0,r+d}c_{1,r}\Big).\end{split}\ee\end{widetext} The fact that they are indeed  zero mode generators results from  the non-trivial commutation relations $[\cT^{(\lambda)}_R,\hat P^{(i)}_d]=\sum\limits_{\lambda'=1}^4\alpha_{\lambda,\lambda',i,R,d}\cT^{(\lambda')}_{R-\frac{d}{2}}$ for $\lambda, i=1,2,3,4$, where $\alpha_{\lambda,\lambda',i,R,d}$ is a coefficient depending on $\lambda,\lambda',i,R,d$.

Simple calculations show that  $\hat p^{a,b}_d$s and $\hat p_d$ in the main paper are essentially equivalent to the above zero mode generators. Indeed, we have
\be
\label{Psubstitution}\begin{split}
&\hat p^{0,0}_d=\hat P^{(2)}_d+d\hat P^{(1)}_d-\hat P^{(3)}_d,\,\,\hat p^{0,1}_d=\hat P^{(1)}_d,\,\,\hat p^{1,0}_d=\hat P^{(4)}_d,\\&\hat p^{1,1}_d=\hat P^{(3)}_d, \,\,\hat p_d=\hat p^{0,0}_d+\hat p^{1,1}_d=\hat P^{(2)}_d+d\hat P^{(1)}_d.
\end{split} \ee
As shown in Eq. \ref{Lie}, $\hat p^{a,b}_d$s form a graded Lie algebra, $[\hat p^{a,b}_k,\hat p^{b',a'}_{k'}]=\delta_{b,b'}\hat p^{a,a'}_{k+k'}-\delta_{a,a'}\hat p^{b',b}_{k+k'}$.
Now if we define $ Q_R^{(1)} $ and $ Q_R^{(4)} $ as  linear combinations of  $ \cT_R^{(1)} $ and $ \cT_R^{(4)} $:\be\begin{split}
& Q_R^{(1)}=\sqrt{\frac{1}{34} \left(17-\sqrt{17}\right)}\cT_R^{(1)}-\sqrt{\frac{1}{34} \left(17+\sqrt{17}\right)}\cT_R^{(4)},\\& Q_R^{(4)}=\sqrt{\frac{1}{34} \left(17+\sqrt{17}\right)}\cT_R^{(1)}+\sqrt{\frac{1}{34} \left(17-\sqrt{17}\right)}\cT_R^{(4)},
\end{split} \ee the zero mode condition (Eq. \ref{zero}) becomes \be\begin{split}\label{zeroB}
&\cT_R^{(\lambda)}\ket{\psi_{\text{zm}}} =0,\quad \mbox{for} \;\; \lambda=2,3,\\& Q_R^{(\lambda')}\ket{\psi_{\text{zm}}} =0,\quad \mbox{for} \;\; \lambda'=1,4.
\end{split} \ee
It is easy to verify that $\hat p^{a,b}_d$ are indeed zero mode generators by virtue of the following commutators:
\begin{subequations}\label{QP}
\be
[Q_R^{(1)},\hat p_d^{0,0}]=2^{1-\frac{d}{2}}\sqrt{\frac{(2R+1)!}{(2R-d+1)!}}\,Q_{R-\frac{d}{2}}^{(1)},
\ee
\be
\begin{split}
[ \cT_R^{(2)},\hat p_d^{0,0}]= &2^{(1-d)/2}\sqrt{\frac{(2R-1)!}{(2R-d+1)!}}\Big(\frac{2d(d-1)}{3}Q_{R-\frac{d}{2}}^{(1)}\\ &+\sqrt{2(2R-d)(2R-d+1)}\cT_{R-\frac{d}{2}}^{(2)} \\
& +d(d-1)Q_{R-\frac{d}{2}}^{(4)}\Big).
\end{split}
\ee\be
[\cT_R^{(3)},\hat p_d^{0,0}]=2^{1-\frac{d}{2}}\sqrt{\frac{(2R)!}{(2R-d)!}}\,\cT_{R-\frac{d}{2}}^{(3)},
\ee
\be
[Q_R^{(4)},\hat p_d^{0,0}]=2^{1-\frac{d}{2}}\sqrt{\frac{(2R+1)!}{(2R-d+1)!}}\,Q_{R-\frac{d}{2}}^{(4)},
\ee
\be
[Q_R^{(1)},\hat p_d^{0,1}]=0,
\ee
\be
\begin{split}
[\cT_R^{(2)},\hat p_d^{0,1}]=& -\frac{2^{(3-d)/2}	}{3}\sqrt{\frac{(2R-1)!}{(2R-d+1)!}}\Big((d-1)Q_{R-\frac{d}{2}}^{(1)}\\
& +\sqrt{2R-d+1}\cT_{R-\frac{d}{2}}^{(3)}+2(d-1)Q_{R-\frac{d}{2}}^{(4)} \Big),
\end{split}
\ee\be
[\cT_R^{(3)},\hat p_d^{0,1}]=2^{1-\frac{d}{2}}\sqrt{\frac{(2R)!}{(2R-d+1)!}}\,Q_{R-\frac{d}{2}}^{(4)},
\ee
\be
[Q_R^{(4)},\hat p_d^{0,1}]=0,
\ee
\be
\begin{split}
[Q_R^{(1)},\hat p_d^{1,0}]=& 2^{-\frac{d}{2}}\sqrt{\frac{(2R+1)!}{(2R-d+1)!}}\Big((d+1) Q_{R-\frac{d}{2}}^{(1)}\\
& +(2R+1)\sqrt{2R-d+1}\cT_{R-\frac{d}{2}}^{(3)} \Big),
\end{split}
\ee\be
\begin{split}
[\cT_R^{(2)},\hat p_d^{1,0}]=& \frac{2^{(1-d)/2}	}{3}\sqrt{\frac{(2R-1)!}{(2R-d+1)!}}\\&\Big((1+d)R(1+2R)Q_{R-\frac{d}{2}}^{(1)}\\ &+3\sqrt{2(2R-d)(2R-d+1)}\cT_{R-\frac{d}{2}}^{(2)}\\
& -R(1+2d-2R)\sqrt{2R-d+1}\cT_{R-\frac{d}{2}}^{(3)}\\
&-2(d+1)R(-2+d-4R)Q_{R-\frac{d}{2}}^{(4)}  \Big),
\end{split}
\ee
\be
\begin{split}
[\cT_R^{(3)},\hat p_d^{1,0}]=& 2^{-1-\frac{d}{2}}\sqrt{\frac{(2R)!}{(2R-d+1)!}}\\&\Big(-2(1+d)(2R+1) Q_{R-\frac{d}{2}}^{(1)}\\ &-3\sqrt{2(2R-d)(2R-d+1)}\cT_{R-\frac{d}{2}}^{(2)} \\
& +2(1+d)\sqrt{2R-d+1}\cT_{R-\frac{d}{2}}^{(3)}\\
&+(d^2-d-4-4R^2-4dR-10R)Q_{R-\frac{d}{2}}^{(4)}\Big),\end{split}\ee
\be\begin{split}[Q_R^{(4)},\hat p_d^{1,0}]=& 2^{-\frac{d}{2}}\sqrt{\frac{(2R+1)!}{(2R-d+1)!}}\Big(-(d+1) Q_{R-\frac{d}{2}}^{(1)}\\
& +\sqrt{2R-d+1}\cT_{R-\frac{d}{2}}^{(3)} \Big),
\end{split}
\ee
\be
\begin{split}
[Q_R^{(1)},\hat p_d^{1,1}]=& 2^{-\frac{d}{2}}\sqrt{\frac{(2R+1)!}{(2R-d+1)!}}\Big(Q_{R-\frac{d}{2}}^{(1)}\\& -Q_{R-\frac{d}{2}}^{(4)}\Big),
\end{split}
\ee
\be
\begin{split}
[\cT_R^{(2)},\hat p_d^{1,1}]=& \frac{2^{(3-d)/2} }{3}\sqrt{\frac{(2R-1)!}{(2R-d+1)!}}\Big(d R Q_{R-\frac{d}{2}}^{(1)}\\
& -R\sqrt{2R-d+1}\cT_{R-\frac{d}{2}}^{(3)}+2 d R Q_{R-\frac{d}{2}}^{(4)} \Big),
\end{split}
\ee
\be
\begin{split}
[\cT_R^{(3)},\hat p_d^{1,1}]=& 2^{-\frac{d}{2}}\sqrt{\frac{(2R)!}{(2R-d+1)!}}\Big(d Q_{R-\frac{d}{2}}^{(1)}\\
& +\sqrt{2R-d+1}\cT_{R-\frac{d}{2}}^{(3)}\\
&+(1+2d+2R)Q_{R-\frac{d}{2}}^{(4)} \Big),
\end{split}
\ee
\be
[Q_R^{(4)},\hat p_d^{1,1}]=2^{1-\frac{d}{2}}\sqrt{\frac{(2R+1)!}{(2R-d+1)!}}\,Q_{R-\frac{d}{2}}^{(4)},
\ee\end{subequations}
Most importantly, the operators appearing on the right-hand sides are always linear combinations of the operators in \Eq{zeroB}, and thus vanish within the zero mode subspace. This ensures that the action of any of the $\hat p^{a,b}_d$ on any zero mode gives a new zero mode.

Now we will prove that the $ \hat e_k $ defined in Eq. (\ref{e}) satisfy the Newton-Girard formula (\ref{ng}). Therefore, these operators are $ k $-body zero mode generators as they can be expressed in terms of the $\hat p_d$ with $ d=1,...k $. As a result, $\hat  S_{\ell}  $ is also a zero mode generator by its definition. To prove the Newton-Girard formula, we can write down $\hat  e_k $ in terms of $\hat  e_{k-1} $, \be \hat e_k =\frac{1}{k}\sum\limits_{n,l}\tilde c_{n, {l} + 1}^*  \hat e_{k-1}\tilde c_{n,l}.\ee Using the commutator $ [\hat e_k,\tilde c_{n,l}]=- \hat e_{k-1}\tilde c_{n, l - 1} $ to move the $ \hat e $ operator all the way to the right of the $ \tilde c  $ operators, one can arrive at the Newton-Girard formula \be\label{ng}\hat e_k=\frac{1}{k}\sum\limits_{d=1}^{k}(-1)^{d-1}\hat p_d \hat  e_{k-d}. \ee
In the same way, one can use $ [\hat e^{a,b}_k,\tilde c_{b,l}]=-\delta_{a,b}\hat e^{a,b}_{k-1} \tilde c_{b, l - 1} $ to obtain a modified Newton-Girard formula\be \hat e^{a,b}_k=\frac{1}{k}\hat p^{a,b}_1\hat e^{a,b}_{k-1}+\frac{\delta_{a,b}}{k}\sum\limits_{d=2}^{k}(-1)^{d-1}\hat p^{a,b}_d \hat e^{a,b}_{k-d}.\ee Consequently, $ \hat e^{a,b}_k $   are also   $ k $-body zero mode generators since they can be expressed in terms  of either $ \hat p^{a,b}_1$ or $ \hat p^{a,b}_d$ with $ d=1,...k $.

With  Eq. \ref{ng} and the above (modified) Newton-Girard  formulas, we immediately see that $\hat S$ and $\hat  e^{a,b}$ are zero mode generators.

\section{$A(r)$ matrix for $n$ LLs}\label{AppA}
Here we generalize the transformation  matrix $A(r)$ for 2 LLs to the case of $ n $ LLs. Its entries are \be\label{An} A(r)_{ij}=\frac{(i+r)!\,i!}{(i-j)!}\frac{1}{\sqrt{(j+r)!j!} }\,,\quad (i\geq j)\,,\ee
and vanish for $i<j$, as well as for $i<-r$, $j<-r$, as obtained straightforwardly by expanding
monomials $\bar z^{i}z^{i+r}$ (Gaussians are omitted) in
disk Landau level wave functions.
The $A(r)_{ij}$ are basically the expansion coefficients, up to $i$-dependent normalization factors that we dropped for simplicity, as they do not affect the properties of the operators defined in the main text in any essential way. (Note that Table \ref{norm} does contain these extra factors, for the special case of the LLL.)
The operators $\tilde{c}^*_{i,r}=\sum_j A(r)_{ij}c^\dagger_{j,r} $ therefore create single-particle states proportional to the monomials $\bar z^{i}z^{i+r}$.

Strictly speaking, $A(r)$ is invertible only for $r\geq 0$. However, we leave understood that at any given $r$, we always work
within the range of $A(r)$ (thus ignoring unphysical indices $i,j<-r$).
With this restriction in mind, we can always invert $A(r)$ to obtain a likewise
 lower-triangular matrix with the following non-zero
 entries: \be\label{InAn} A^{-1}(r)_{ij}=(-1)^{i+j}\frac{\sqrt{(i+r)!i!}}{(i-j)!(j+r)!{j!}} \,,\quad (i\geq j)\,.\ee
For definiteness, we still take $A^{-1}(r)_{ij}=0$ for $i$ or $j$ being $<-r$.

\begin{acknowledgments}
Work in Florida is funded in part by DOE, Office of BES through Grant No. DE-SC0002140, and performed at the National High Magnetic Field Laboratory, which is supported by NSF Cooperative Agreements No. DMR-1157490 and DMR-1644779, and the State of Florida. AS would like to thank J.K. Jain for insightful discussions.
\end{acknowledgments}

\bibliography{RF}\end{document}